\documentclass{aa}

\usepackage{amsmath}
\usepackage{siunitx}
\usepackage{xfrac}
\usepackage{natbib}
\usepackage{xcolor}
\usepackage{graphicx}
\usepackage{booktabs}
\usepackage{csquotes}
\usepackage[varg]{txfonts}
\makeatletter
  \renewcommand*\aa@pageof{, page \thepage{} of \pageref*{LastPage}}
\makeatother
\usepackage{hyperref}
\hypersetup{pdfpagemode = {UseNone},
            pdftitle = {Cool Cooling Title!},
            pdfcreator = {\LaTeX},
            pdfproducer = {pdfeTeX-0.\the\pdftexversion\pdftexrevision},
            pdfauthor = {Wilhelm Kley, Daniel Thun, Anna Penzlin},
            pdfsubject = {},
            pdfview = {FitH},
            pdfstartview = {FitH},
            colorlinks = {true},
            linkcolor = [rgb]{0,0.35,0.7},
            citecolor = [rgb]{0,0.35,0.7},
            filecolor = [rgb]{0.61,0,0},
            urlcolor = [rgb]{0.61,0,0},
           }

\usepackage{bm}
\renewcommand{\vec}[1]{\bm{\mathrm{#1}}}

\newcommand*\dd[1]{\,\mathrm{d}#1}

\newcommand{\Prt}[1]{\frac{\partial}{\partial #1}}

\newcommand*{\vnabla}{\vec{\nabla}}

\colorlet{darkgreen}{green!60!black}

\newcommand{\beq}{\begin{equation}}
\newcommand{\eeq}{\end{equation}}

\begin{document}

\title{Circumbinary discs with radiative cooling and embedded planets}

\author{Wilhelm Kley \and
        Daniel Thun  \and
        Anna~B.~T. Penzlin }

\institute{
Institut f\"ur Astronomie und Astrophysik, Universität T\"ubingen,
Auf der Morgenstelle 10, D-72076 T\"ubingen, Germany\\
\email{\{wilhelm.kley, daniel.thun, anna.penzlin\}@uni-tuebingen.de}\\
}

\date{}

\abstract
{As of today ten circumbinary planets orbiting solar type main sequence stars have been discovered.
Nearly all of them orbit around the central binary very closely to the region of instability where it is
difficult to form them in situ.
Hence, it is assumed that they formed further out and migrated to their observed position, which will be determined
by binary, disc and planet properties.
}
{We extend previous studies to a more realistic thermal disc structure and
determine what parameter influence the final parking location of a planet around a binary star.
}
{We perform two-dimensional numerical simulations of viscous accretion discs around a central binary that include
viscous heating and radiative cooling from the disc surfaces. We vary the binary eccentricity as well as disc viscosity and mass.
}
{Concerning the disc evolution we find that it can take well over \num{100000} binary orbits until an equilibrium state is reached.
As seen previously, we find that the central cavity opened by the binary becomes eccentric and precesses slowly in a prograde
sense. Embedded planets migrate to the inner edge of the disc. In cases of lower disc viscosity they migrate further
in maintaining a circular orbit,
while for high viscosity they are parked further out on an eccentric orbit.
}
{Discs around binary stars are eccentric and precess very slowly around the binary.
The final location of an embedded planet is linked to its ability to open a gap in the disc.
Gap opening planets separate inner from outer disc, preventing eccentricity excitation in
the latter and making it more circular. This allows embedded planets to migrate closer to the binary, in agreement with the
observations. The necessary condition for gap opening and the final planet position depend on the planet mass and disc viscosity.
}


\maketitle

\section{Introduction}\label{sec:intro}
Circumbinary discs are accretions discs that surround two central objects such as
a system of binary black holes or binary stars.
Here, we are dealing with protostellar discs that orbit around a binary system consisting of two
protostars. We are interested in the dynamical evolution of such circumbinary discs and
in addition we study the evolution of planets in them.

About ten planets orbiting main sequence stellar binaries in a Tatooine like manner have been discovered until
today, all of them by the Kepler space telescope. The systems show planetary transits for both stars,
hence they are all very flat, an overview of their orbital parameter is given in
\citet{2018MNRAS.480.3800H} and references therein.
This suggests a coplanar configuration of binary orbit and protostellar disc
during the formation phase of the planetary systems. As a consequence, and to simplify the simulations,
two-dimensional hydrodynamical studies of infinitesimally thin discs around binary stars are
a good approximation. In this spirit, several simulations of circumbinary discs with and without embedded planets
have been performed
\citep{2002A&A...387..550G,2007A&A...472..993P,2008ApJ...672...83M}, some with explicit reference
to the observed Kepler systems \citep{2013A&A...556A.134P,2014A&A...564A..72K,2015A&A...581A..20K,2017MNRAS.469.4504M,
2018A&A...616A..47T}.

Due to angular momentum transfer from the binary to the disc a central cavity develops where the density is
strongly reduced. The size of this induced gap depends on
disc parameter such as viscosity and pressure, and binary parameter such as orbital eccentricity and mass ratio
\citep{1994ApJ...421..651A}.
Longterm high resolutions simulations have shown that the formed gaps around circumbinary stars feature a complicated dynamical
evolution \citep{2017MNRAS.466.1170M,2017A&A...604A.102T,2018A&A...616A..47T} with details depending on binary and disc
parameter.
In all cases the gap and the inner disc region turns eccentric and precesses slowly in a prograde manner around the binary star.
Surprisingly, the highest gap eccentricity develops for circular binaries. Upon increasing the binary eccentricity,
$e_\mathrm{bin}$, the gap first become more circular and then more eccentric again above a critical value
of $e_\mathrm{bin} \approx 0.16$ \citep{2017A&A...604A.102T}. Varying the binary mass ratio shows that for more
equal mass binaries the precession period increases for all binary eccentricities.
Results of sequences of simulations for different binary parameter are shown in \citet{2018A&A...616A..47T}, where
the overall variation of the gap orbital dynamics is compared to test particle trajectories around the binary.

In spite of the formation of the gap the central binary can still accrete material from the disc that flows into the
gap and is added onto central circumstellar discs.
This mass flow within the cavity around the stars has been studied in a several simulations
\citep{1994ApJ...421..651A,2002A&A...387..550G,2010ApJ...708..485H,2011MNRAS.413.2679D}.
The gas dynamics near the central objects is relevant in
another example of circumbinary discs that refers to a central binary black hole system which is orbited by a disc,
for example after a merger process of two galaxies in the early Universe.
Of particular interest in this case is the evolution of the orbital elements of the binary
as it accretes mass and angular momentum from the disc and exchanges angular momentum with the disc
via gravitational torques.
While earlier studies found a decrease of the orbit of the binary due to the angular momentum loss to the disc \citep{2009MNRAS.393.1423C},
recent work suggests that an expansion of the binary orbit due to mass
and accompanying angular momentum accretion is also possible
\citep{2017MNRAS.466.1170M,2019ApJ...871...84M}.
As we are interested here in the overall longterm evolution and structure of the disc we do not study
the central region in this work and leave this interesting topic to future investigations.

The detected circumbinary planets all orbit the binary star close to the dynamical instability limit \citep{2018ApJ...856..150Q}.
The strong dynamical action of the binary makes an in situ formation at this close vicinity very difficult and it is
typically assumed that the planets formed further out in the disc and then migrated to their present location.
Indeed, simulations of embedded planets show that they migrate inwards due to the planet disc interaction
and then are 'parked' at the inner disc edge, which is the outer rim of the gap 
\citep{2007A&A...472..993P,2008A&A...483..633P,2013A&A...556A.134P,2014A&A...564A..72K}.
Its final location will be determined by the size and structure of the central cavity region which depends,
as mentioned above, on binary as well as disc parameter.

Most of the studies on circumbinary disc dynamics work with a simplified thermodynamical model using a locally isothermal
disc structure where the temperature (or sound speed) in the disc is a given function of radius. As the few exceptions
\citep{2002A&A...387..550G,2014A&A...564A..72K} suffer from low numerical resolution or too short time span, we perform here
a detailed study of the disc structure including viscous heating and radiative cooling from the disc surfaces.
Starting from two baseline models using Kepler-16 and Kepler-38 parameter we
vary binary eccentricity as well as disc mass and viscosity
and compute the disc evolution until an equilibrium state is reached. Finally, we embed planets into the disc and follow their
evolution. We show that the final position of the planet is determined by its ability to open its own gap in the disc.

\begin{table}
    \centering
    \begin{tabular}{ccccccc}
        \midrule\midrule
        Kepler & $M_A$ & $M_B$ & $q_\mathrm{bin}$ &
        $a_\mathrm{bin}$ & $e_\mathrm{bin}$ &
        $T_\mathrm{bin}$ \\
               & $[M_\sun]$ & $[M_\sun]$ &  &
        $[\mathrm{au}]$ &   &
        $[\mathrm{d}]$ \\
        \midrule
       16 &  0.67 & 0.20 & 0.29 & 0.22 & 0.16 & 41.0 \\
       38 &  0.95 & 0.25 & 0.26 & 0.15 & 0.10 & 18.8 \\
        \midrule
    \end{tabular}
    \caption{Orbital parameter of the stellar binaries in the Kepler-16 and Kepler-38 systems.
    Here, $M_A$ and $M_B$ denote the stellar masses of the binary, and its 
    mass ratio is defined as $q_\mathrm{bin} = M_B/M_A$. The orbital elements are the semi-major axis, $a_\mathrm{bin}$,
    the eccentricity, $e_\mathrm{bin}$, and the period, $T_\mathrm{bin}$.
     References: \citet{2011Sci...333.1602D,2012ApJ...758...87O}
    \label{tab:kepler}
   }
\end{table}

The paper is organised as follows: In Sect.~\ref{sec:setup} we describe the assumptions of our physical and numerical model.
This is followed in Sect.~\ref{sec:disc_structure} by a description of the structure of radiative circumbinary discs, for various
binary eccentricities and different values for the disc viscosity.
The evolution of an embedded planet in a radiative disc using Kepler-38 parameter is presented in Sect.~\ref{sec:planet}.
Our results are finally discussed in Sect.~\ref{sec:discussion} and summarised in Sect.~\ref{sec:summary}.

\section{Model setup}\label{sec:setup}
We extend our previous locally isothermal circumbinary disc models \citep{2017A&A...604A.102T,2018A&A...616A..47T}
to a more realistic equation of state, study the impact on the
disc structure and dynamics, and follow the orbital evolution of an embedded planet.
We perform two-dimensional (2D) hydrodynamical, viscous simulations including an energy equation.
To account for energy losses and prevent the disc from overheating by viscous
dissipation we consider also radiative cooling from the disc surfaces.
We study the evolution of two sample systems, Kepler-16 and Kepler-38, their parameters are summarised in
Table~\ref{tab:kepler}.

\begin{figure}
    \centering
    \resizebox{\hsize}{!}{\includegraphics[trim={3.0cm 23.5cm 5.0cm 3.0cm},clip]{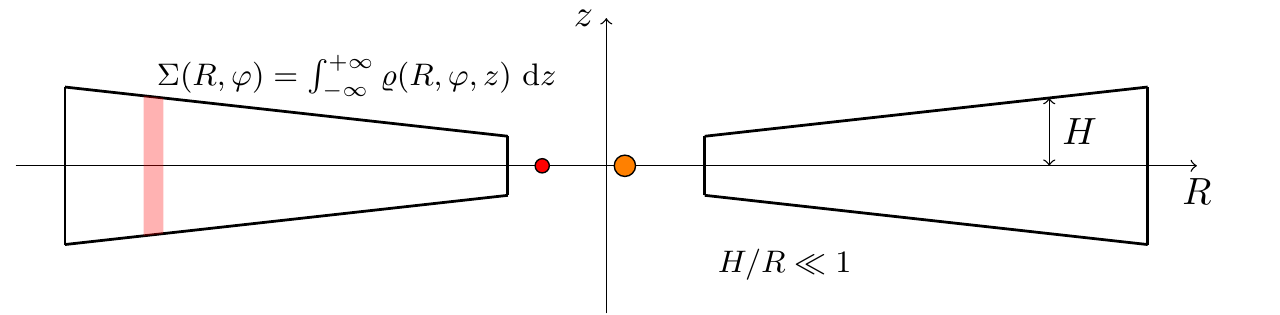}}
    \caption{Sketch of our disc model with the averaging process indicated for
    the three-dimensional density $\varrho$. The orange circle marks the primary
    and the red circle the secondary star for the Kepler-38 system, where the sizes are
    proportional to the stellar masses.
    The origin of the coordinate system lies at the barycentre of the binary.}
    \label{img:disc_sketch}
\end{figure}
\subsection{Physical model}
To model the circumbinary discs we use cylindrical coordinates ($R$,
$\varphi$, $z$) centred on the centre of mass (COM) of the binary with the disc lying
in the $z=0$-plane. The observed Kepler systems with a circumbinary
planet are nearly coplanar which justifies to work in 2D. We
therefore assume that the vertical extent of the disc, $H$, is small compared to
the radial extension, and use the vertically averaged
hydrodynamical equations, see Figure~\ref{img:disc_sketch}.
To be concise we quote briefly only the main equations that are solved and will refer
to our previous studies for more details.

The continuity equation is given by
\begin{equation}\label{eq:continuity}
    \Prt{t}\Sigma + \vnabla \cdot \left(\Sigma \vec{u} \right) = 0\,,
\end{equation}
with the surface density $\Sigma = \int_{-\infty}^{\infty} \varrho \dd z$ \
(see Fig.~\ref{img:disc_sketch}) and
the 2D velocity $\vec{u} = (u_R, u_\varphi)^\mathrm{T}$.
The vertically averaged equation of momentum conservation is given by
\begin{equation}\label{eq:momentum_conservation}
    \Prt{t}\left(\Sigma \vec{u} \right) + \vnabla \cdot \left(\Sigma \vec{u}
    \otimes \vec{u} - \vec{\Pi} \right) = -\vnabla P  + \Sigma \, \vec{g}\,,
\end{equation}
with the vertically integrated pressure, $P = \int_{-\infty}^{\infty} p \dd z$,
and viscous stress tensor $\vec{\Pi}$. The individual components in cylindrical coordinates
are given in \citet{1999MNRAS.303..696K} or \citet{2002A&A...387..605M}, where we use here a vanishing bulk viscosity.
To model the turbulent shear viscosity in the disc we use the standard
$\alpha$-disc model by \citet{1973A&A....24..337S}. In this approximation, the
kinematic viscosity coefficient can be written for thin discs as
\beq
  \label{eq:nu}
  \nu = \alpha c_s H \,,
\eeq
with a parameter $\alpha < 1$, the adiabatic sound speed of the gas $c_s$, and the vertical pressure scale height (disc half thickness)
$H$.

The gravitational acceleration, $\vec{g}$, acting on each cell due to the binary and the planet is
given in detail described in \citet{2018A&A...616A..47T} and we do not go into details here.
It is given by the sum of the gravitational forces of the two binary stars, a possible planet and
takes into account a possible acceleration of the COM of the binary.
The gravitational acceleration contains a smoothing length
$\epsilon H$ to avoid singularities and to account for the correct
treatment of the three-dimensional gravity in our 2D setup
\citep{2012A&A...541A.123M}. In all simulations we use $\epsilon = 0.6$ \citep{2002A&A...387..605M}
and evaluate the pressure scale height, $H$, at the location of the cell.
Assuming a vertical hydrostatic equilibrium and an isothermal disc in the $z$-direction the disc
height is given by
\begin{equation}\label{eq:disc_height}
    H = \sqrt{\frac{P}{\Sigma}} \, \left[ \sum_\ell \frac{G M_\ell}{\left| \vec{R} -
    \vec{R}_\ell \right|^3}  \right]^{-1/2}\,,
\end{equation}
where $\vec{R}$ denotes the position of the grid cell and $\vec{R}_\ell$ the position of a star or planet.
The summation in eq.~\eqref{eq:disc_height} runs over all gravitating objects (stars and planets).
In the approximation $\vec{R}_\ell \rightarrow 0$, i.e. for large distances from the binary, this simplifies to
\begin{equation} \label{eq:disc_height_simplified}
   H =  \sqrt{ \frac{P}{\Sigma} \, \frac{ |\vec{R}|^3}{G M_\mathrm{bin}} }  \,,
\end{equation}
where the mass of the binary is $M_\mathrm{bin} = M_\mathrm{A} + M_\mathrm{B}$.
We carried out test simulations to compare the sophisticated disc
height~\eqref{eq:disc_height} to the simple
one~\eqref{eq:disc_height_simplified}, and did not find any significant differences. Therefore, we use
in all our simulations eq.~\eqref{eq:disc_height_simplified} to calculate the
height of the disc.

The vertically averaged equation of energy conservation reads
\begin{equation}\label{eq:energy_conservation}
    \Prt{t}\left(\Sigma e\right) + \vnabla \cdot \left[\left(\Sigma e + P
     - \vec{\Pi} \right) \vec{u} \right] = \left( \Sigma \vec{u} \right) \cdot
     \vec{g} - Q_-  \,.
\end{equation}
The total specific energy $e$ is related to the pressure by
\begin{equation}
    \Sigma e = \frac{P}{\gamma - 1} + \frac{1}{2} \Sigma \vec{u}^2 \,,
\end{equation}
with the adiabatic index $\gamma$. In all our simulations we use a value of
$\gamma = 1.4$.  
The pressure is related to the midplane temperature $T$ of the disc
\begin{equation}
\label{eq:temp}
    P = \frac{k_\mathrm{B}}{\mu m_\mathrm{H}} \Sigma T \,,
\end{equation}
with the Boltzmann constant $k_\mathrm{B}$, the mass of the hydrogen atom
$m_\mathrm{H}$ and the mean molecular mass of the gas $\mu$ (we use $\mu =
2.35$). The adiabatic sound speed of the gas is then given by
\begin{equation}
    c_s = \sqrt{\gamma \frac{P}{\Sigma}}\, .
\end{equation}

\begin{figure}
    \centering
    \resizebox{\hsize}{!}{\includegraphics{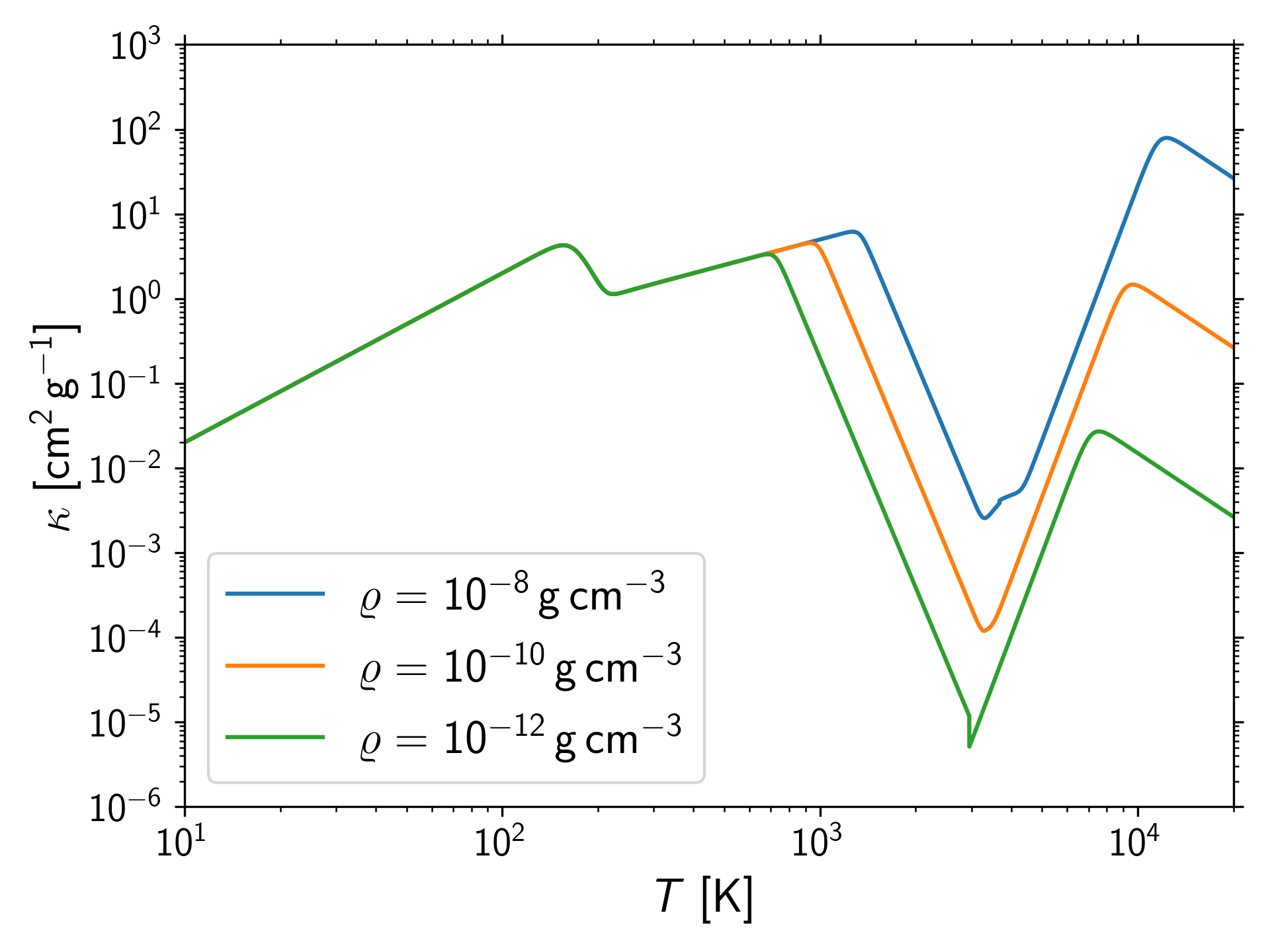}}
    \caption{Variation of the Rosseland mean opacity with temperature for three
    different densities $\varrho$.}
    \label{img:rosseland_mean_opacity}
\end{figure}
To model the cooling term $Q_-$ we follow~\citet{2003ApJ...599..548D} and
\citet{2012A&A...539A..18M} and only consider the cooling effect of radiation in
the vertical direction and neglect radiation transport in the disc midplane.
This is a good assumption if the vertical extent of the disc is small compared
to the discs' radial extent. The cooling term models the energy loss by radiation
from the upper and lower disc surface and can be written as
\begin{equation}\label{eq:cooling_term}
    Q_- = 2\sigma_\text{R} \frac{T^4}{\tau_\mathrm{eff}},
\end{equation}
with the Stefan-Boltzmann constant $\sigma_\text{R}$. Since radiation can escape
on both sides of the disc the factor of two in eq.~\eqref{eq:cooling_term}
is necessary. To calculate the effective optical depth $\tau_\mathrm{eff}$ we
follow an approach by \citet{1990ApJ...351..632H} as given in \citet{2012A&A...539A..18M}.
\begin{equation}\label{eq:tau_eff}
    \tau_\mathrm{eff} = \frac{3}{8}\tau + \frac{\sqrt{3}}{4} + \frac{1}{4 \tau +
    \tau_\mathrm{min}} \,.
\end{equation}
For the mean optical depth we use the following approximation
\begin{equation}\label{eq:optical_depth}
    \tau = \int \varrho \kappa \dd{z} \approx  \varrho_0 \kappa H \, = \, \frac{c}{\sqrt{2\pi}} \, \Sigma \kappa \,,
\end{equation}
where  $\varrho_0$ denotes the midplane gas density and
$c = \sfrac{1}{2}$ is a constant to account for the drop of opacity with vertical height
in our two dimensional simulations \citep{2012A&A...539A..18M}.
To calculate the Rosseland mean opacity $\kappa$ we
assume a power-law dependence on temperature and density
\citep{1985prpl.conf..981L}
\begin{equation}\label{eq:mean_opacity}
    \kappa = \kappa_0 \, \varrho_0^a \, T^b\,,
\end{equation}
where $\varrho_0$ and $T$ refer to the midplane gas density and temperature.
The values of $\kappa_0$, $a$, and $b$ for different opacity regimes can be found
in a table given in \citet{2012A&A...539A..18M}.
Figure~\ref{img:rosseland_mean_opacity} shows the opacity variation with
temperature for three different densities. To calculate the opacity we use a
function which smoothly transitions from one regime to another. We apply a minimum optical depth of
$\tau_\mathrm{min} = 0.1$ to account for the very optically thin regions of the disc.

In our simulations we do not consider the self-gravity of the disc as we are primarily interested
in the end phase of the planet migration process when the disc mass is already reduced. For simulations of
discs with self-gravity and embedded planets, see \citet{2017MNRAS.465.4735M,2017MNRAS.469.4504M}.
We also do not consider stellar irradiation and will discuss this in Sect.~\ref{subsec:irradiation}.

\subsection{Initial conditions}
For the initial disc parameters we follow \citet{2018A&A...616A..47T} and use an
initial surface density profile given by
\begin{equation}\label{eq:init_sigma}
    \Sigma(t=0) = f_\mathrm{gap} \Sigma_\mathrm{ref} \left(
    \frac{R}{a_\mathrm{bin}} \right)^{-1.5} \,.
\end{equation}
The function $f_\mathrm{gap}$ models the expected cavity created by the binary
\citep{1994ApJ...421..651A, 2002A&A...387..550G}
\begin{equation}
    f_\mathrm{gap} = \left[1+\exp{\left(-\frac{R-R_\mathrm{gap}}
                                              {\Delta R}\right)} \right]^{-1} \,,
\end{equation}
with $R_\mathrm{gap} = 2.5\,a_\mathrm{bin}$ and $\Delta R =
0.1\,R_\mathrm{gap}$. The reference density $\Sigma_\mathrm{ref}$ is calculated
according to the disc mass $M_\mathrm{disc}$ inside the computational domain,
and typically we use a total initial disc mass of $0.01 M_\mathrm{bin}$
(for details see \citet[Sec. 2.2]{2018A&A...616A..47T}).
For the pressure profile we assume a disc with an initial constant aspect ratio
$h (t=0) = H/R = 0.05$ and calculate the initial pressure from
eq.~\eqref{eq:disc_height_simplified}
\begin{equation}\label{eq:init_prs}
    P(t=0) = h(t=0)^2 \frac{G M_\mathrm{bin}}{R} \Sigma(t=0)\,.
\end{equation}
The initial radial velocity is set to zero $u_R(t= 0) = 0$ and the initial
azimuthal velocity is set to the local Keplerian velocity with respect to the
COM of the binary $u_\varphi(t=0) = \sqrt{G M_\mathrm{bin}/R}$.

\subsection{Numerical considerations}\label{ssec:numerics}
All simulations were performed by a modified version of \textsc{Pluto} 4.2
\citep{2007ApJS..170..228M} which runs on GPUs and has in addition a N-body
module to simulate the motion of planets in binary stars. Details can be
found in \citet{2018A&A...616A..47T}
The new implementation of the cooling terms into the code is described in Appendix~\ref{app:cooling}.
An overview of the numerical options for the hydrodynamical solver used here is
given in \citet[Sec. 3.1]{2017A&A...604A.102T}.
The numerical setup is identical to the setup described in \citet[Sec.
2.3]{2018A&A...616A..47T}. We use a logarithmically expanding grid in the radial
direction from $R_\mathrm{min} = 1\,a_\mathrm{bin}$ to $R_\mathrm{max} =
40\,a_\mathrm{bin}$ and a uniform grid from $0$ to $2\pi$ in the azimuthal
direction. In all our simulations this domain is covered by a grid with fixed resolution of
$N_r  \times N_\varphi = 684 \times 584$ cells.
At $R_\mathrm{min}$ a zero-gradient boundary condition ($\partial/\partial R =
0$) is used for surface density, radial velocity, angular velocity
$\Omega_\varphi = u_\varphi/R$ and pressure. Additionally, no inflow into the
computational domain is allowed at $R_\mathrm{min}$. At $R_\mathrm{max}$ we set
the azimuthal velocity to the Keplerian velocity. For surface density and radial
velocity we use a wave damping boundary condition, where $u_R$ is damped towards zero and
$\Sigma$ to the initial value.
\citep{2006MNRAS.370..529D,2018A&A...616A..47T}. For the pressure we use again a
zero gradient boundary condition. In the azimuthal direction we use
periodic boundary conditions.
For stability reasons we use a global density floor of
\begin{equation}
    \Sigma_\mathrm{floor} = 10^{-6} \Sigma_\mathrm{ref}\,,
\end{equation}
for a discussion of this value see \citet{2018A&A...616A..47T}. To enhance numerical stability
we also use a temperature floor of $T_\mathrm{floor} = \SI{3}{K}$ and a temperature ceiling of
$T_\mathrm{ceiling} = \SI{3000}{K}$. The lower value prevents material from too low temperatures while
the upper ceiling is relevant in the initial phase of the simulations in order to prevent too large temperatures
in the inner gap region.

\begin{figure}
    \centering
    \resizebox{\hsize}{!}{\includegraphics{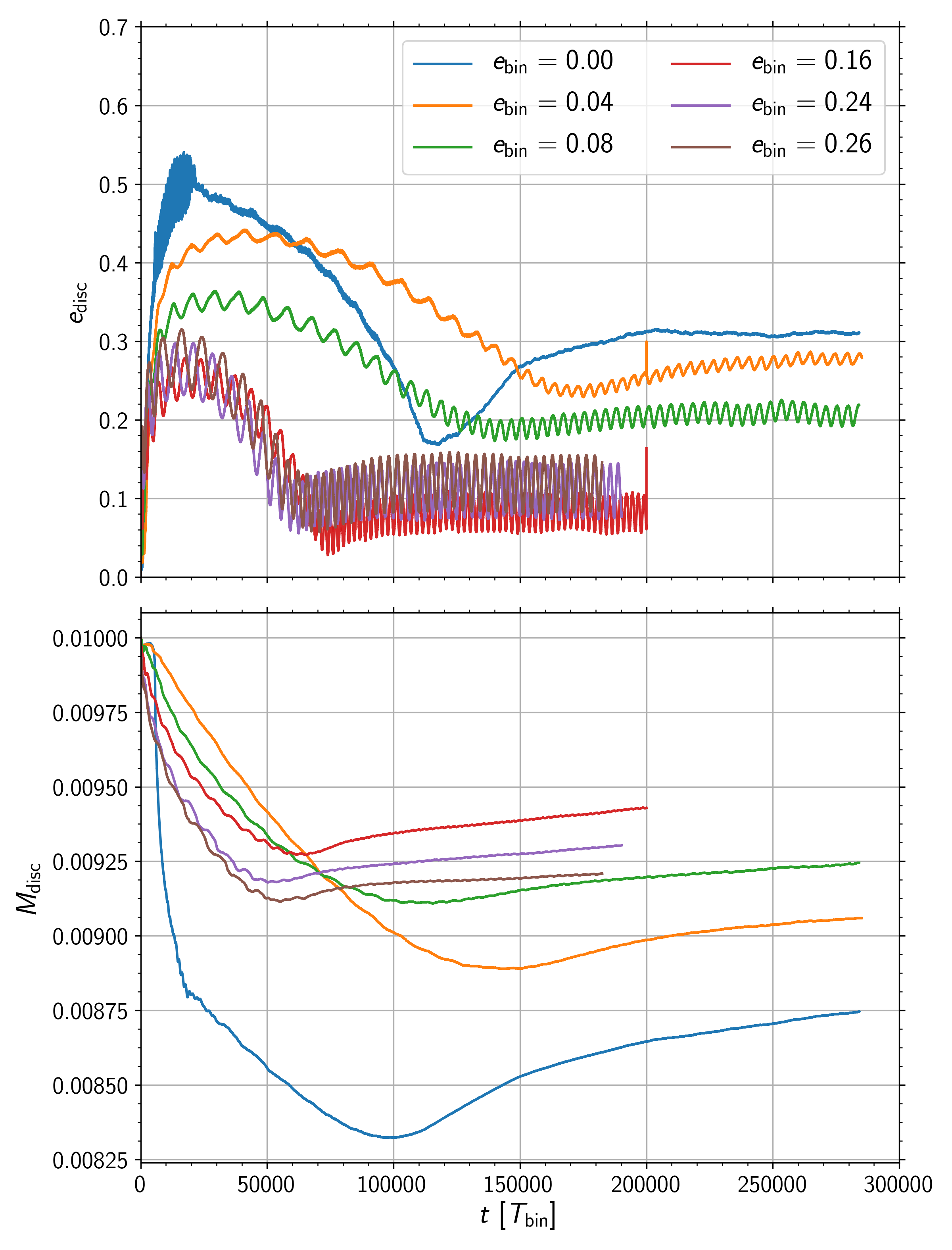}}
    \caption{The time evolution of the disc eccentricity and mass for fixed
    $q_\mathrm{bin}=0.29$ (as in Kepler-16) and various binary eccentricities over time.
    The top panel shows the disc eccentricity, while the bottom
    panel shows the total mass in the disc in units of the central binary mass.}
    \label{img:Kep16-disc_evolution}
\end{figure}

\section{Disc structure}\label{sec:disc_structure}
In this Section we focus on the disc structure created by the gravitational
interaction between the binary and the disc. We are interested in the size and
eccentricity as well as the precession period of the gap as a function of binary eccentricity
and physical disc properties.
First, we describe a sequence of models for a fixed binary mass ratio $q_\mathrm{bin}=0.29$
as in the Kepler-16 system, and vary the binary eccentricity $e_\mathrm{bin}$.
As the thermodynamics of the disc is determined by the disc mass and viscosity we present
subsequently models using the fixed binary parameter of Kepler-38 for different
$M_\mathrm{disc}$ and viscosity $\alpha$. The orbital parameter of Kepler-16 \& 38 are listed in Tab.~\ref{tab:kepler}.
\begin{figure}
    \centering
    \resizebox{\hsize}{!}{\includegraphics{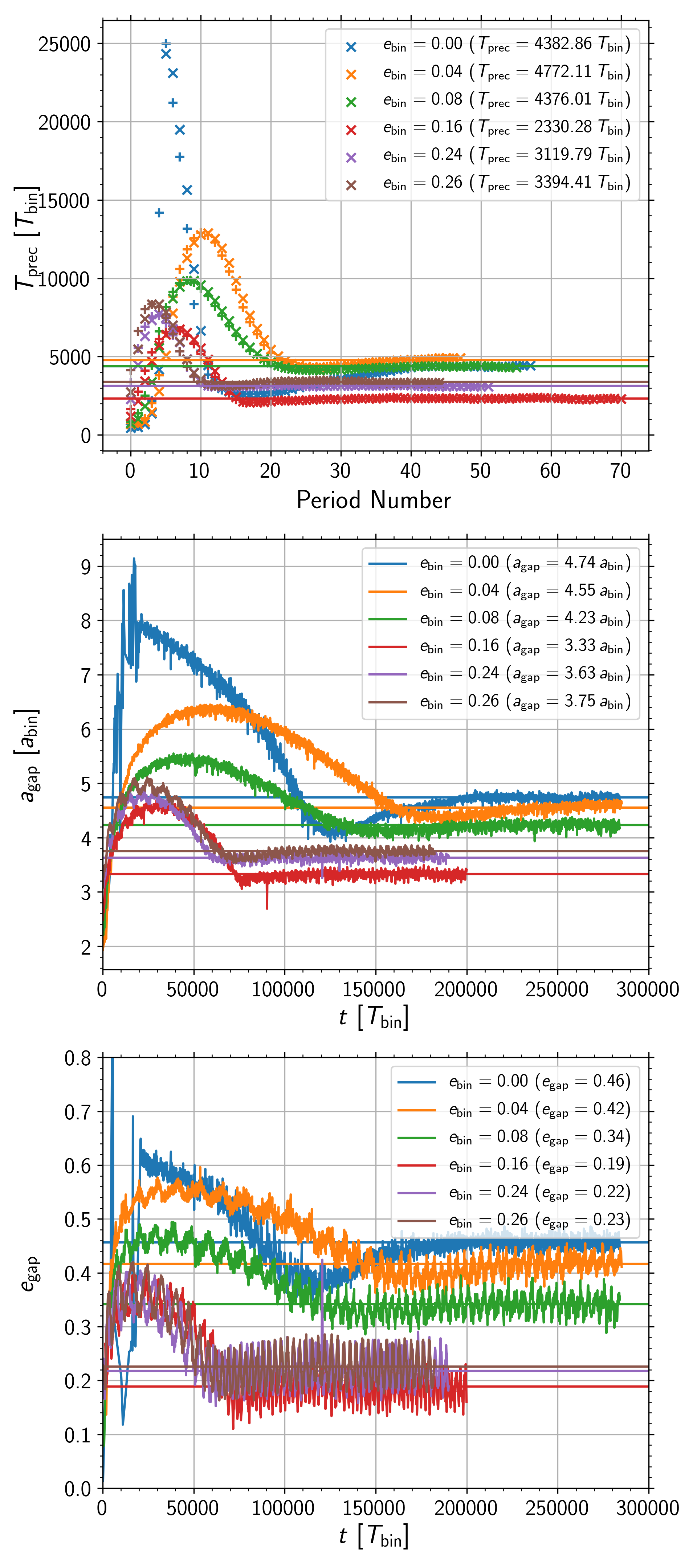}}
    \caption{The evolution of the gap properties for fixed $q_\mathrm{bin}=0.29$
    and various binary eccentricities.  The top panel shows the
    precession period of the gap as a function of period number. The bottom two
    panels show the gap's semi-major axis and eccentricity.
    The horizontal lines and labels indicate the final values.}
    \label{img:Kep16-disc_props_ecc}
\end{figure}

\subsection{Models for different binary eccentricity}\label{ssec:kep16_sequence}
\label{seubsec:Kepler16-sequence}
For the model sequence presented in this section we use a fixed disc viscosity
$\alpha = 0.001$ and an initial disc mass of $0.01 M_\mathrm{bin}$. The total mass
and mass ratio of the binary are those of the Kepler-16 system as quoted in
Tab.~\ref{tab:kepler}. We run a sequence of models for various binary eccentricities
running from $e_\mathrm{bin} =0$ to $e_\mathrm{bin} =0.28$.

In Fig.~\ref{img:Kep16-disc_evolution} we display the evolution of the disc mass and eccentricity
    for various $e_\mathrm{bin}$ over time.
    The top panel shows the disc eccentricity, where an averaging has been performed
    over the inner disc area out to a radius of $R = 8.0\,a_\mathrm{bin}$ \citep{2017A&A...604A.102T}.
    The bottom panel shows the time evolution of the total disc mass in units of the central binary mass.
    Clearly noticeable is the very long time needed until the disc reaches a quasi-equilibrium state.
    However, this timescale depends on the binary eccentricity.
    For $e_\mathrm{bin}$ larger than about 0.16 the discs settle in less than $\num{100000}\,T_\mathrm{bin}$
    while for lower $e_\mathrm{bin}$ the equilibration time well exceeds  $\num{150000}\,T_\mathrm{bin}$.

    The discs around more circular binary stars with $e_\mathrm{bin} < 0.1$ show a larger disc eccentricity, $e_\mathrm{disc}$.
    The minimum of $e_\mathrm{disc}$ is reached around $e_\mathrm{bin} = 0.16$ with a small increase for larger $e_\mathrm{bin}$.
    The mass loss from the disc through the inner boundary is moderate overall, during the $\num{200000}\,T_\mathrm{bin}$ the total mass
    loss is at most 15\%. In the very early phase, the mass loss is larger for the more eccentric binaries because the secondary
    periodically approaches the inner edge of the disc 'pulling out' some material from it.
    Upon the strong increase of the disc eccentricity, particularly for the $e_\mathrm{bin}=0$ model, the mass loss increases
    suddenly. In the long run all the models settle to a stationary state, as the density  at the outer radius of the disc is
    held fixed at its initial value. During this adjustment phase the disc mass can also increase as the fixed density 
    at $R_\mathrm{max}$ serves as a mass reservoir for the inner disc.

In Fig.~\ref{img:Kep16-disc_props_ecc} we display the evolution of the gap geometry for the same models
    over time. The properties of the gap are here obtained by fitting an ellipse to the inner gap edge,
    where the density has reached about half of the maximum density near the inner rim of the disc.
    The exact procedure is described in \citet{2017A&A...604A.102T}, and examples of such fitted ellipses are
    shown in Fig.~\ref{img:Kep38-2d_sigma_disc_physics} below.
    The time evolution of the obtained semi-major axis,
    eccentricity and precession period of the ellipses are shown in Fig.~\ref{img:Kep16-disc_props_ecc}.
    The horizontal lines in the figure and the labels indicate the values measured in the final phase of the simulations.
    The long process of finding the equilibrium is seen clearly in the lower two panels. In all models the gap size and
    eccentricity increase initially very strongly, reach their maxima between $\num{20000}$ and about $\num{60000}\,T_\mathrm{bin}$,
    and then settle slowly towards equilibrium. The final gap eccentricities (bottom panel) are largest for the circular binary,
    reach the minimum at $e_\mathrm{bin} = 0.16$ and then increase again.
    The variation of the gap semi-major axis (middle panel) follows the same trend:
    the largest gap sizes are seen for low $e_\mathrm{bin}$ and the smallest for $e_\mathrm{bin} = 0.16$.
    As seen from the top panel in the figure, the precession rate of the ellipse becomes very long for the discs with large gaps,
    and reach over $\num{10000}\,T_\mathrm{bin}$ for the smaller $e_\mathrm{bin}$ cases. Eventually, the precession period settles
    to values between about 2300 and 4770\,$T_\mathrm{bin}$ for all models.
   This disc evolution is different to that of the locally isothermal discs presented in \citet{2018A&A...616A..47T}.
   There we did not observe the initial strong increase in the hole size and eccentricity. We will comment on this difference in
   Sect.~\ref{subsec:local_iso} below.

  The time evolution for the disc eccentricity in Figs.~\ref{img:Kep16-disc_evolution} and \ref{img:Kep16-disc_props_ecc}
   displays oscillatory behaviour for $e_\mathrm{disc}$ or $e_\mathrm{gap}$, respectively. The oscillations are a secular effect,
   they have been seen already in \citet{2017A&A...604A.102T} and discussed in \citet{2017MNRAS.466.1170M}.
   The oscillation period is identical to the precession period of the disc, i.e. the time for one $360^\circ$ turn of the eccentric disc.
   Similar to single test particle orbits, the period of the oscillations becomes longer for larger cavity sizes
   (corresponding to larger distances of the particle) and shorter for smaller cavities \citep{2017A&A...604A.102T}.
   This explains why the oscillations are wider in the early phase with the extended gap size,
   and are narrower in the relaxed equilibrium state.

\begin{figure}[t]
    \centering
    \resizebox{\hsize}{!}{\includegraphics{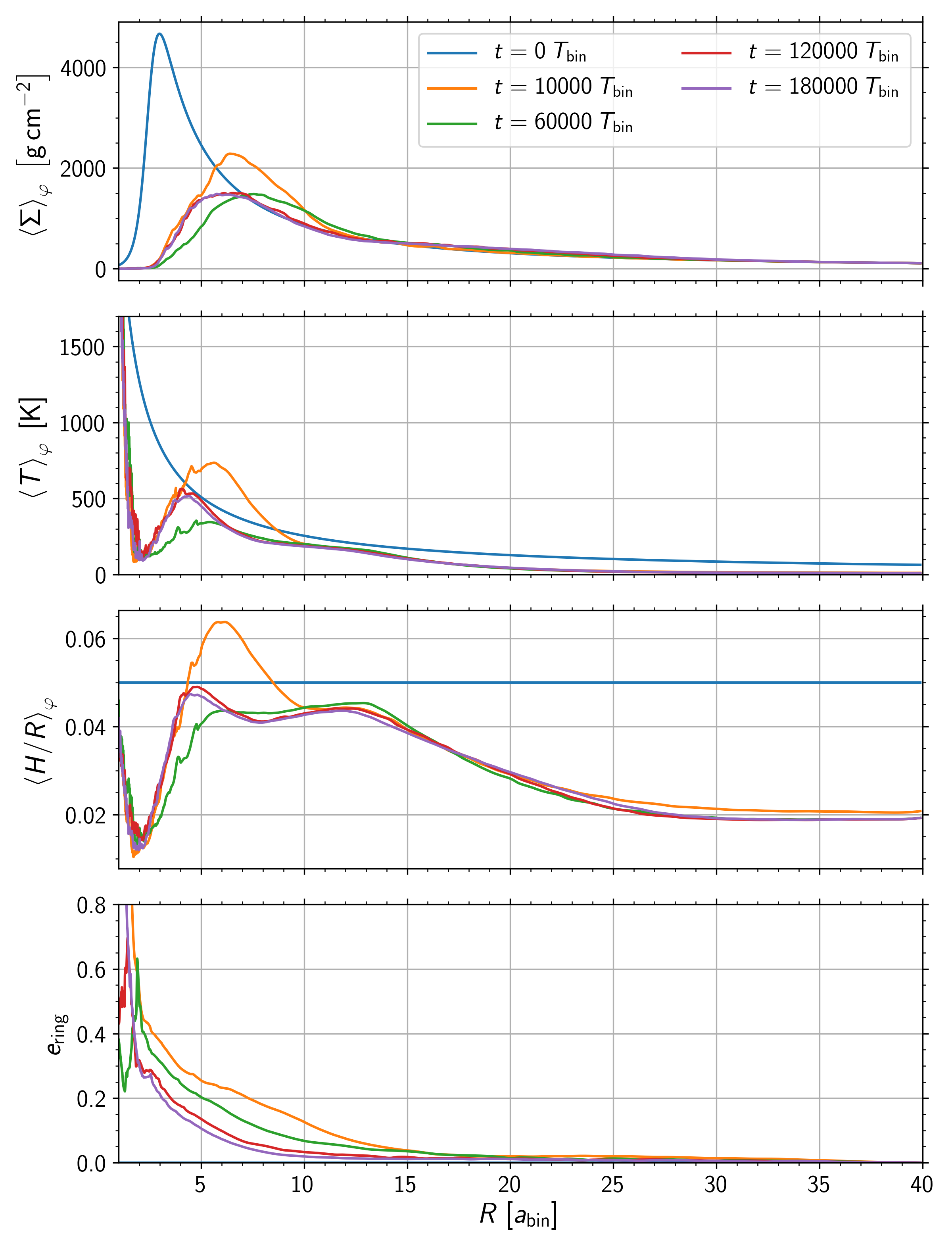}} \\
    \caption{Azimuthally averaged radial disc profiles for $e_\mathrm{bin}=0.16$ at various evolutionary times.
    Displayed are the surface density, the midplane temperature, the disc thickness and the
    disc eccentricity.
    }
    \label{img:Kep16-radial_plots_016}
\end{figure}

The azimuthally averaged radial profile of various disc quantities are shown in
Fig.~\ref{img:Kep16-radial_plots_016} for the Kepler-16 case with
$e_\mathrm{bin}=0.16$ at five evolutionary times. The initial model is shown as
reference.  The next two models at $\num{10000}$ and
$\num{60000}\,T_\mathrm{bin}$ demonstrate the early adjustment phase, and
the two final models at $\num{120000}$ and $\num{180000}\,T_\mathrm{bin}$
that the disc has indeed reached its equilibrium state. While
initially the disc experiences a phase with a large eccentric gap, in the final
state, the surface density is identical to the initial state beyond about $7 a_\mathrm{bin}$.
The disc's averaged inner edge (about half value of $\Sigma_\mathrm{peak}$)
always lies around $4 a_\mathrm{bin}$. Overall, the radiative disc is cooler than the initial setup with
a reduced thickness. In the disc's inner regions inside of
about $15 a_\mathrm{bin}$ the thickness is about $H/R = 0.04$ while in the outer
regions the thickness reduces to $H/R = 0.02$.  Within the gap the minimum
thickness is again about $H/R = 0.01 - 0.02$. In the very inner parts of the disc near
the inner boundary where the density is very close to the density floor, the
temperature increases strongly, but still remains below the initial value
corresponding to $H/R = 0.05$. This increase in temperature is a numerical
artefact and is discussed in more detail in Appendix~\ref{app:cooling}. Due to
the very small density and pressure this hot gas does not have any dynamical
influence on its environment. In the equilibrium state the disc's eccentricity
(bottom panel) reaches about 0.3 at the inner disc edge (at $3 a_\mathrm{bin}$) and is nearly
circular beyond $15 a_\mathrm{bin}$.

\begin{figure}[t]
    \centering
    \resizebox{\hsize}{!}{\includegraphics{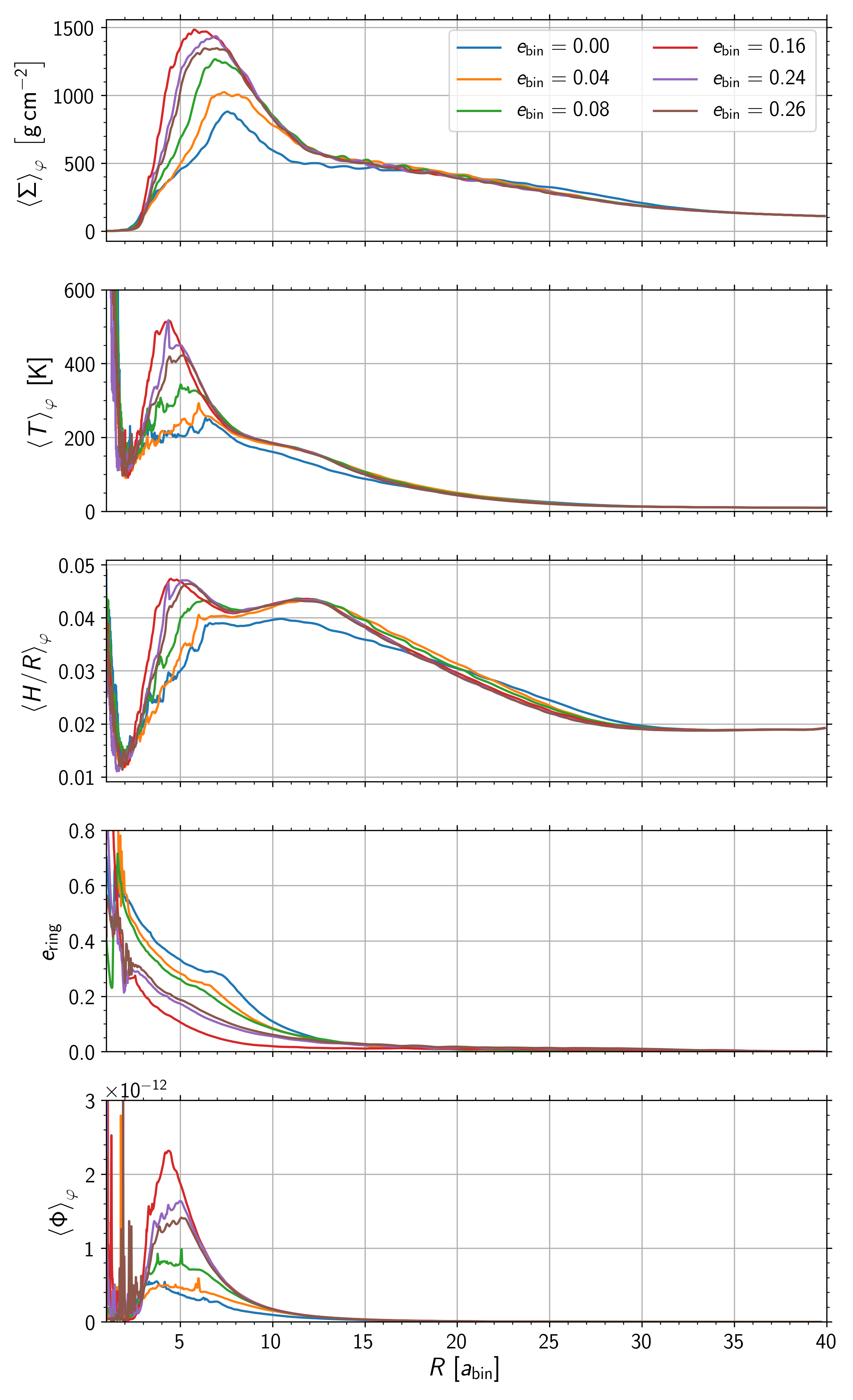}} \\
    \caption{Azimuthally averaged radial disc profiles in equilibrium for different
    $e_\mathrm{bin}$ after $\num{180000}$ binary orbits.
    Displayed are the surface density, the midplane temperature, the disc
    thickness, the disc eccentricity, and in code units the dissipation
    function~\eqref{eq:visc_dissipation}.}
    \label{img:Kep16-radial_plots_final}
\end{figure}

In Fig.~\ref{img:Kep16-radial_plots_final} we display the azimuthally averaged
radial disc profiles for different $e_\mathrm{bin}$ after $\num{180000}$ binary
orbits. At this time all models have reached the final equilibrium
configuration.
As could be already inferred from the calculated gap semi-major axis, in the
models with smaller $e_\mathrm{bin}$ the maximum of the surface density lies
further out in the disc and the peak value is very small.  The turning point
model with the smallest inner hole, for $e_\mathrm{bin}=0.16$, has the largest
density spike which lies closest to the binary.  Due to the higher density the
disc is optically thicker and the midplane temperature is higher for the models
with larger $e_\mathrm{bin}$. The temperature maxima for all models lie within
the range 4 to 7 $a_\mathrm{bin}$, well inside of the peak density. 
For the models with the larger binary eccentricities, the
temperature peak coincides with the maximum of viscous dissipation at around
$4-5\,a_\mathrm{bin}$, as indicated in the bottom panel in
Fig.~\ref{img:Kep16-radial_plots_final}.
For the dissipation function we considered only the the
$R\varphi$-component of the viscous stress tensor
\begin{equation}\label{eq:visc_dissipation}
    \Phi = \frac{9}{4} \Sigma \nu \left(R \frac{\partial \Omega}{\partial R}
    \right)^2 \,.
\end{equation}
For the small binary eccentricities the temperature maximum is shifted to larger radii due to additional contributions by 
shock heating which is given by the term $-P  \, \nabla \cdot \vec{u}$.
The disc eccentricity follows the same trend as the gap
eccentricity. For smaller $e_\mathrm{bin}$ the discs are more eccentric over a
larger radial domain.

\begin{figure}
    \centering
    \resizebox{\hsize}{!}{\includegraphics{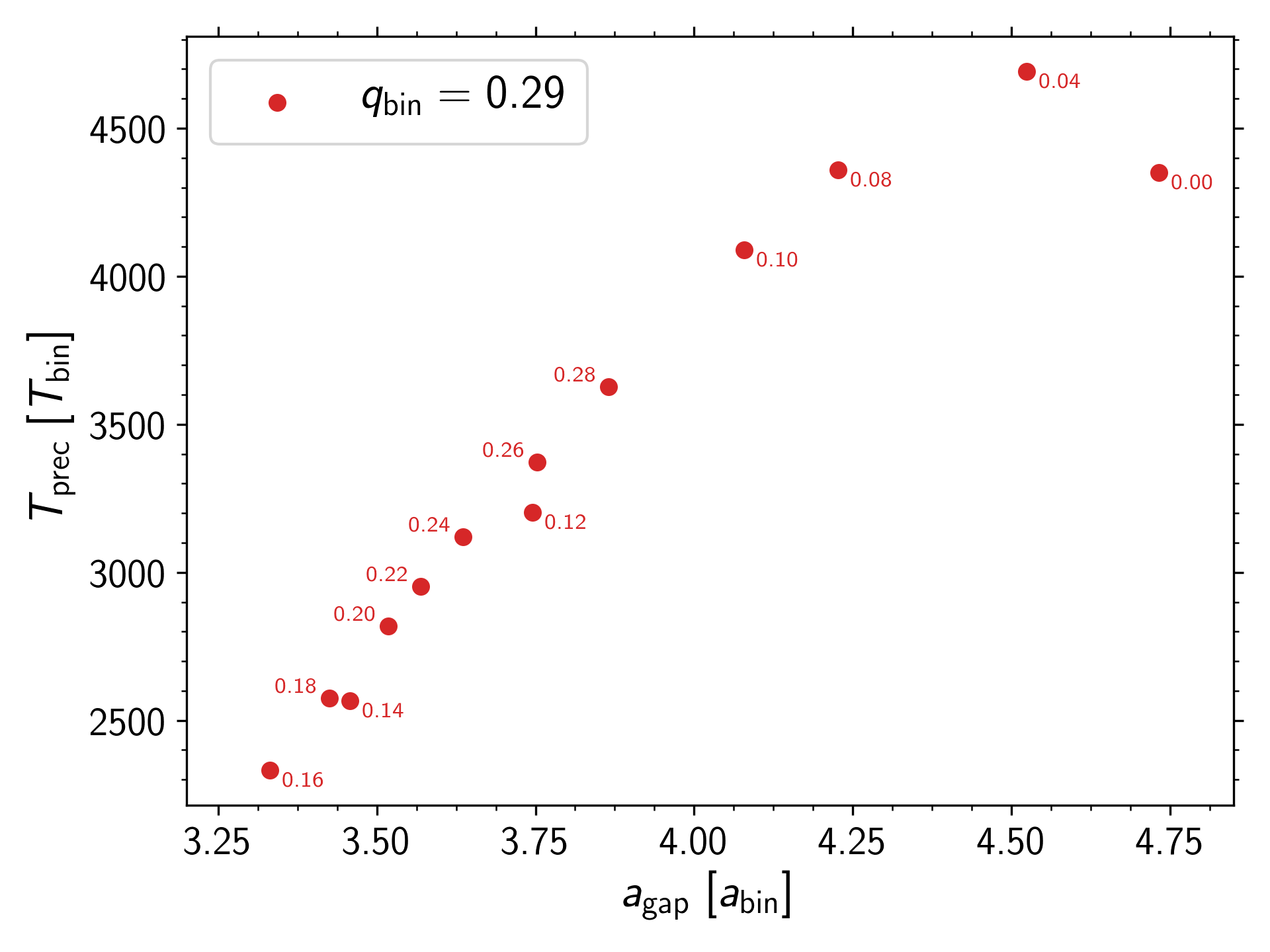}} \\
    \caption{Gap precession period against gap size in equilibrium after $\num{180000}$ binary orbits.
    All the models have the same $q_\mathrm{bin}=0.29$ and different $e_\mathrm{bin}$ as indicated by the labels.
    }
    \label{img:Kep16-Tprec_vs_agap}
\end{figure}
In Fig.~\ref{img:Kep16-Tprec_vs_agap} we display the gap precession period
against gap size for models with identical $q_\mathrm{bin}=0.29$ and varying
$e_\mathrm{bin}$.  The models for small $e_\mathrm{bin}$ have large holes and
long precession periods.  Increasing $e_\mathrm{bin}$ reduces the gap size and
ellipticity and makes the disc precess faster.  The turning point at which this
trend reverses again, occurs at a critical value of $e_\mathrm{bin} = 0.16$.
This value is identical to the critical point obtained for locally isothermal
discs (with $H/R=0.5$) and with a viscosity of $\alpha = 0.01$, as shown in the
bifurcation diagram in \citet{2018A&A...616A..47T}. For this radiative disc the
upper and lower branch lie very close to each other.  Scaling of the precession
period with the free particle orbits, as done in \citet{2018A&A...616A..47T},
would separate the branches vertically.

\begin{figure}
    \centering
    \resizebox{\hsize}{!}{\includegraphics{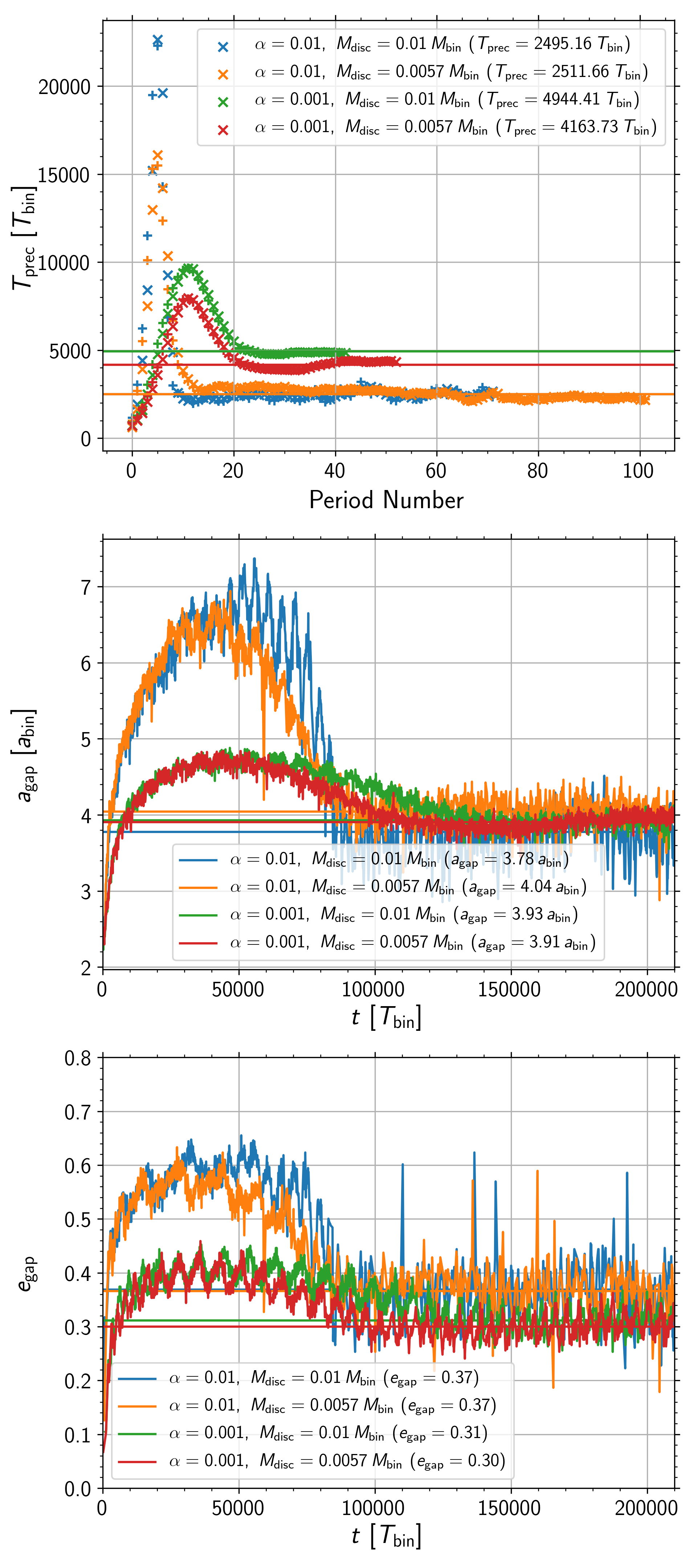}}
    \caption{Precession period, semi-major axis, and eccentricity of the disc
    gap vs. time, for discs with two different disc masses and viscosities.
    For the central binary we use the parameter of Kepler-38.
    The horizontal lines indicate time averages starting at
    \num{120000} binary orbits, the time when all discs reached a quasi-steady
    state.}
    \label{img:Kep38-disc_props}
\end{figure}
%
%
\subsection{Models for different disc masses and viscosities}
\label{ssec:Kepler38-models}
For radiative models the disc evolution will be influenced by the total disc mass $M_\mathrm{disc}$ and the viscosity $\alpha$
of the disc as these alter the disc temperature and hence change the dynamics.
To estimate the impact that these parameters have on the disc structure we calculated models with different disc masses
and viscosity.
For the basic binary system we chose here {\bf Kepler-38}, with the stellar masses and binary eccentricity
given in Tab.~\ref{tab:kepler}.

To cover various disc physics we performed altogether four simulations with two different values for the viscosity,
$\alpha = 0.001$ and $0.01$, and two different initial disc
masses, $M_\mathrm{disc} = 0.01 M_\mathrm{bin}$ and $0.0057 M_\mathrm{bin}$.
Fig~\ref{img:Kep38-disc_props} summarises the evolution of the dynamical gap properties in the 4 cases.
As noticed already for the previous model sequence using Kepler-16 parameter here all 4 simulations initially produce a large inner gap
which becomes very eccentric with peak values reached after \num{40000} to \num{60000} binary orbits.
After this, the inner gap shrinks and becomes more circular and settles to an equilibrium after about
\num{100000}\,$T_\mathrm{bin}$.
\begin{figure}
    \centering
    \resizebox{\hsize}{!}{\includegraphics{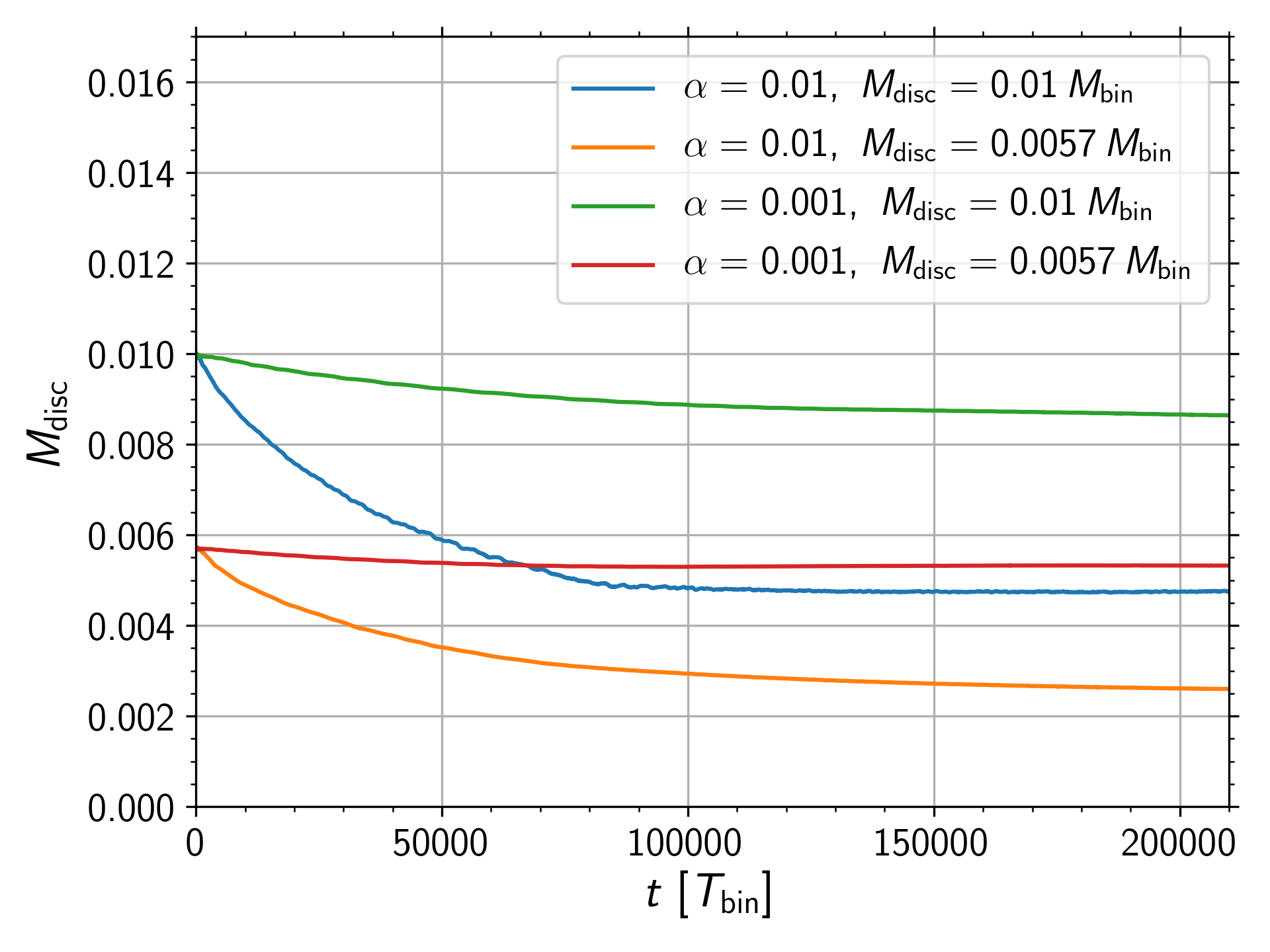}}
    \caption{The disc mass a function of time for models with different viscosity and disc mass.
    }
    \label{img:Kep38-disc_mass_disc_physics}
\end{figure}
\begin{figure}
    \centering
    \resizebox{\hsize}{!}{\includegraphics{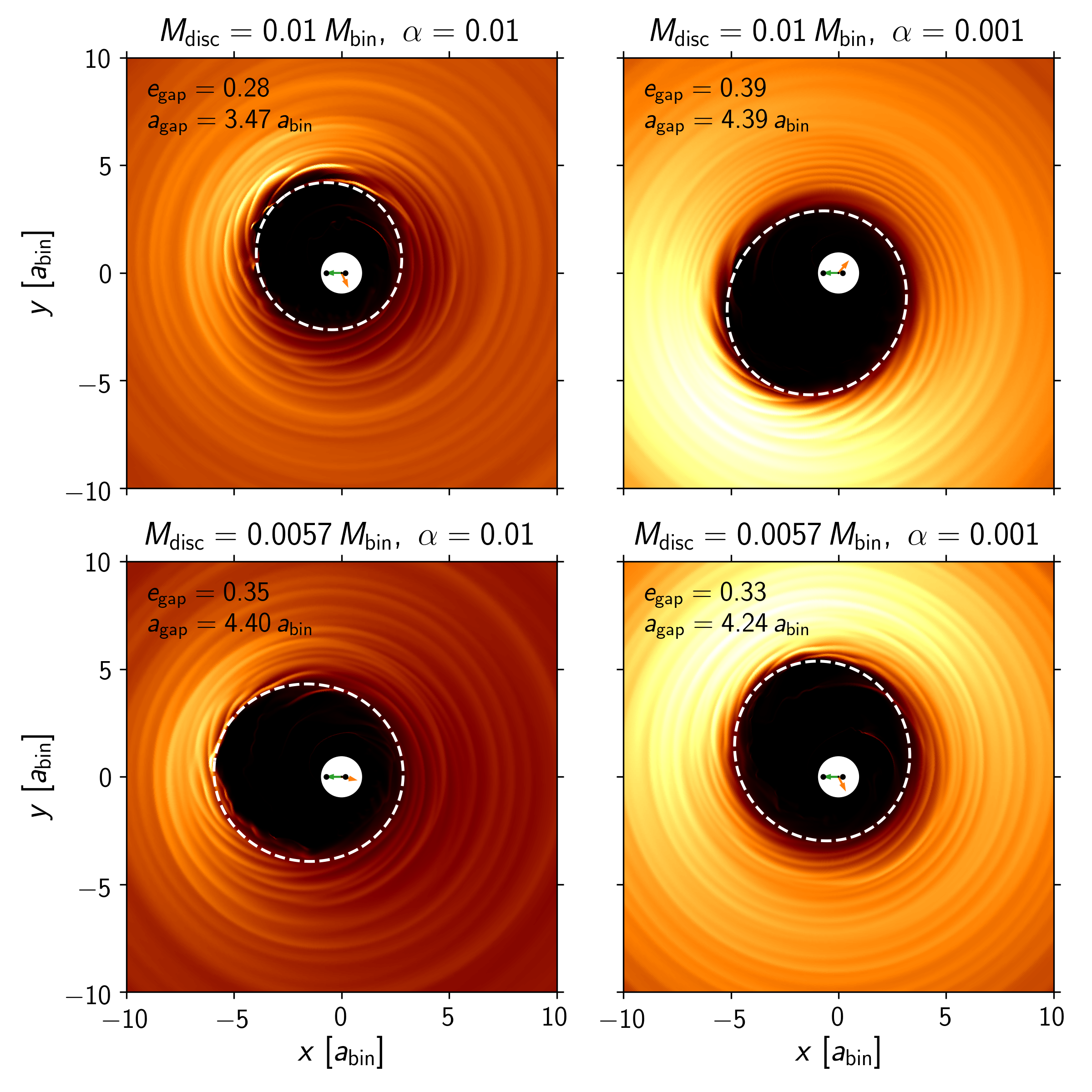}}
    \caption{Two-dimensional density distribution of the inner disc region at a time $t= \num{88000}$.
    The dashed white lines and labels correspond to the approximate sizes and shapes of the inner holes.
    The green arrow points towards the pericentre of the binary and the orange arrow towards the
    pericentre of the inner gap.
    }
    \label{img:Kep38-2d_sigma_disc_physics}
\end{figure}
Discs with high viscosity ($\alpha =0.01$, blue and orange curves in Fig.~\ref{img:Kep38-disc_props})
initially develop larger, and more eccentric gaps with precession periods above \num{15000} $T_\mathrm{bin}$.
Reducing the viscosity by an order of magnitude ($\alpha =0.001$, red and green
curves in Fig.~\ref{img:Kep38-disc_props}) does not produce so large and eccentric gaps in the beginning.
Both, the gap size and eccentricity are near maximum approximately 30\% smaller compared to the high viscosity case.
The precession period of these low viscosity discs is still long and about \num{8000}-\num{10000} binary orbits in the early phase.
During the subsequent evolution all disc models settle to a dynamically similar final state after about \num{100000}\,$T_\mathrm{bin}$,
with gap semi-major axes of $a_\mathrm{gap} = 4 a_\mathrm{bin}$ and  $e_\mathrm{gap}$ between 0.3 and 0.4, depending on $\alpha$.

The evolution of the disc mass for the models is quoted in Fig.~\ref{img:Kep38-disc_mass_disc_physics}.
The two models with the high viscosity, $\alpha = 0.01$, loose mass at a higher rate which is due to the higher viscosity and to the larger
eccentricity of the disc which causes material to be lost through the inner boundary.
After the quasi-equilibrium states have been reached the mass loss from the disc into the inner hole becomes comparable for all cases.
Compared to locally isothermal models with the same viscosity (as presented in \citet{2018A&A...616A..47T})
the radiative models tend to lose more mass.
This could be due to the increased disc eccentricity in the radiative case.
Highly eccentric orbits in the inner part of the disc come very close to
the open inner boundary and gas on these orbits can easily flow through the open
inner boundary.
In contrast to the disc mass evolution presented in Fig.~\ref{img:Kep16-disc_evolution}
where we saw a slight mass increase in the end, we see here a continuous drop in disc mass.
The overall evolution of the total disc mass in the domain is difficult to predict a priori. It depends on the initial
mass in the disc which is not the same for the different cases. The models for Kepler-38
contain initially a higher disc mass ($0.012 M_\odot$) compared to the Kepler-16 models ($0.00089 M_\odot$) 
such that they loose more mass to reach equilibrium.

The two-dimensional density distribution of the central region at a time $t= \num{88000}$ is displayed in
Fig.~\ref{img:Kep38-2d_sigma_disc_physics}. For two models with the smaller viscosity (left panels)
the surface density is lower than for the other two due to larger mass loss from the disc.
The dashed white line corresponds to an elliptical fit to the approximate size and shape of the inner hole.
The method of calculation of the approximate ellipse is described in detail in \citet{2017A&A...604A.102T}.
As mentioned above, all four models end up in a dynamically comparable configuration.

\begin{figure}
    \centering
    \resizebox{\hsize}{!}{\includegraphics{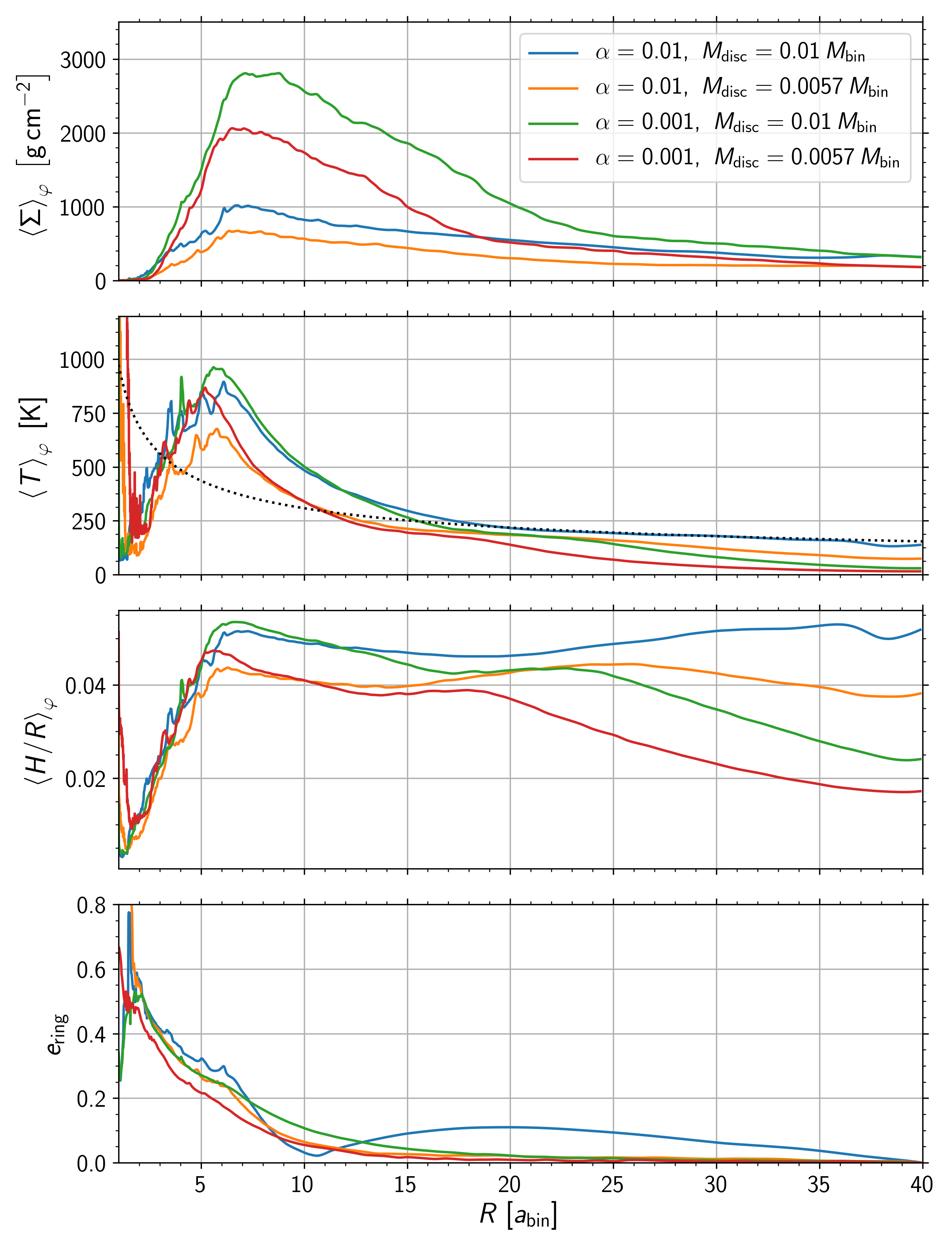}}
    \caption{
    Azimuthally averaged radial disc profiles for the different
    models of the Kepler-38 like system after $\num{120000}$ binary orbits.
    Displayed are the surface density, the midplane temperature, the disc thickness and the
    disc eccentricity. The black dotted line in the 2nd panel represents the irradiation temperature of a
    sun-like star.
    }
    \label{img:Kep38-az_avg_disc_physics}
\end{figure}
The azimuthally averaged radial disc profiles are displayed in Fig.~\ref{img:Kep38-az_avg_disc_physics}
after $\num{120000}$ binary orbits when all models have reached equilibrium.
The two low viscosity models are shown in red and green, and the two high viscosity models in blue and orange.
Due to the increased mass loss through the open inner boundary for high viscosity models the
surface density (top panel) is reduced, while the lower $\alpha = 0.001$ models
show a more pronounced density maximum.
The second panel in Fig.~\ref{img:Kep38-az_avg_disc_physics} shows the
midplane temperature of the disc.
The temperature peaks near the inner rim of the disc just inside of the density maximum due to viscous dissipation.
As shown in the third panel of Fig.~\ref{img:Kep38-az_avg_disc_physics}, the models with the higher initial disc mass
show a final temperature above the starting configuration which corresponds to $H/R = 0.05$, while the models with lower mass lie below it.
Inside the gap, where the density rapidly drops, the cooling is very effective leading to
very low temperatures. This behaviour is in agreement with~\citet{2014A&A...564A..72K}.
In the very inner regions, close to $R_\mathrm{min}$ we see again a strong increase
in temperature which is due to numerically inaccurate cooling (see Appendix~\ref{app:cooling},
where we describe a possible solution).
This feature has no dynamical impact due to the very low densities.
In the outer part of the disc the behaviour of the temperature depends on the viscosity. For
discs with high viscosity the viscous heating is more effective than the cooling
and the temperature is slightly increased above the low viscosity models.
We will discuss the issue of irradiation (black dotted line in the 2nd panel) below in
Sect.~\ref{subsec:irradiation}.
\begin{table}
    \centering
    \begin{tabular}{cccc}
        \midrule\midrule
        $m_p\;[M_\mathrm{Jup}]$ & $a_p [\mathrm{au}]$ & $e_p$  &
        $P_\mathrm{p}\;[\mathrm{d}]$ \\
        \midrule
        0.38 & 0.47 & $<0.03$  & 105 \\
        \midrule
    \end{tabular}
    \caption{Orbital parameter of planet in the Kepler-38 system.
     Reference: \citet{2012ApJ...758...87O}.
    \label{tab:planetkepler}}
\end{table}

Noticeable is the larger disc eccentricity of the model with high viscosity and disc mass, in particular in the outer regions of the
disc. We do not have a full explanation for this behaviour but we believe that it is
related to the different history of the model. Due to the larger disc mass, the 
temperature is higher than for the model with the same viscosity and lower mass (in orange). This makes sound waves propagate faster through the
disc, increasing the interaction with the outer boundary at $R_\mathrm{max}$. In the corresponding temperature and $H/R$ distributions
there is indeed some variation seen near the outer boundary which supports this reasoning.

\begin{figure}
    \centering
    \resizebox{\hsize}{!}{\includegraphics{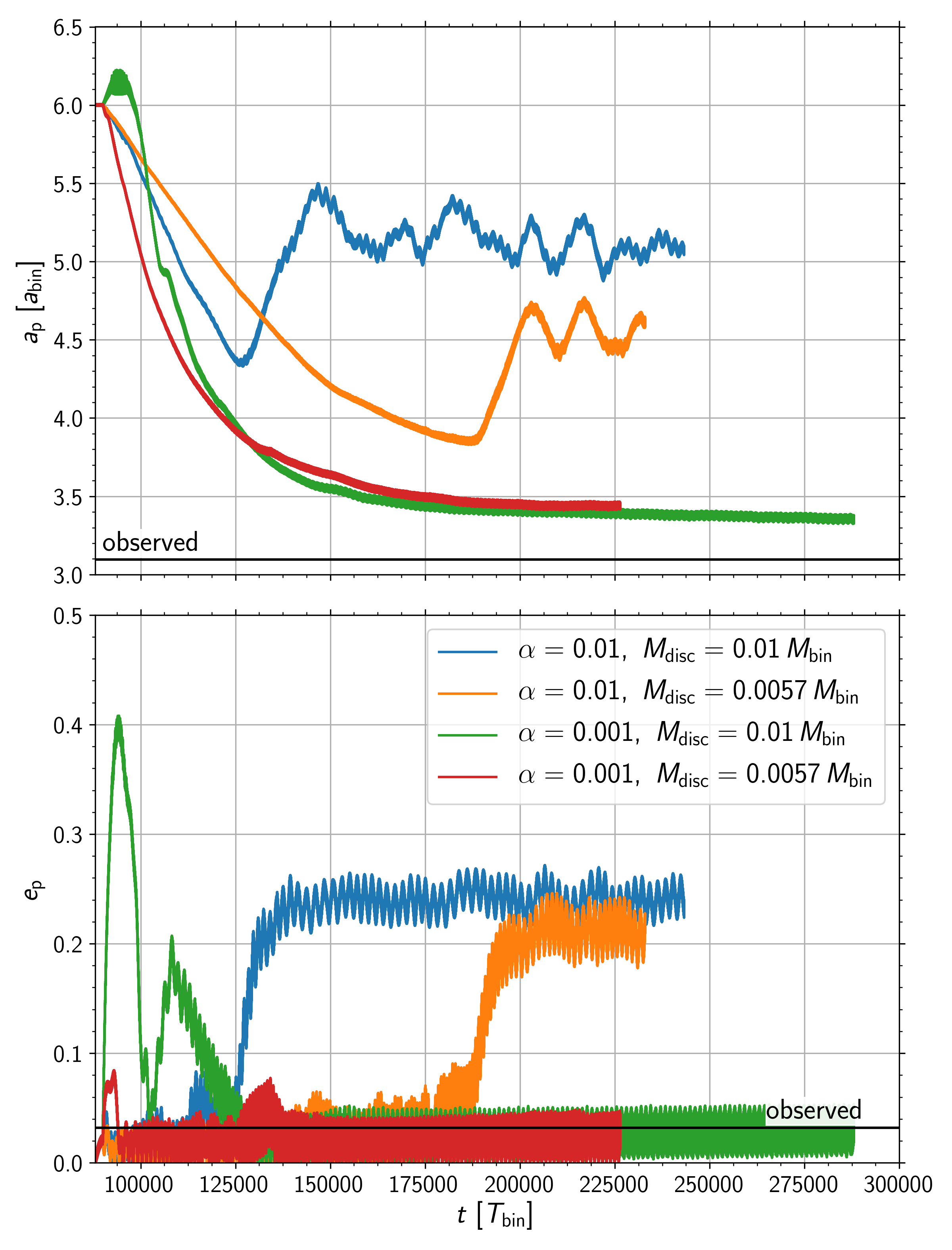}}
    \caption{The evolution of the semi-major axis and eccentricity of an embedded planet in discs
    of different masses and viscosities. The planet is inserted at a time \num{88000} $T_\mathrm{bin}$
    on a circular orbit with $a_\mathrm{p} = 6 a_\mathrm{bin}$ and
    held fixed for the first 2000 $T_\mathrm{bin}$.
    }
    \label{img:Kep38-Planet-orbital_elements_ap60}
\end{figure}
\begin{figure}
    \centering
    \resizebox{\hsize}{!}{\includegraphics{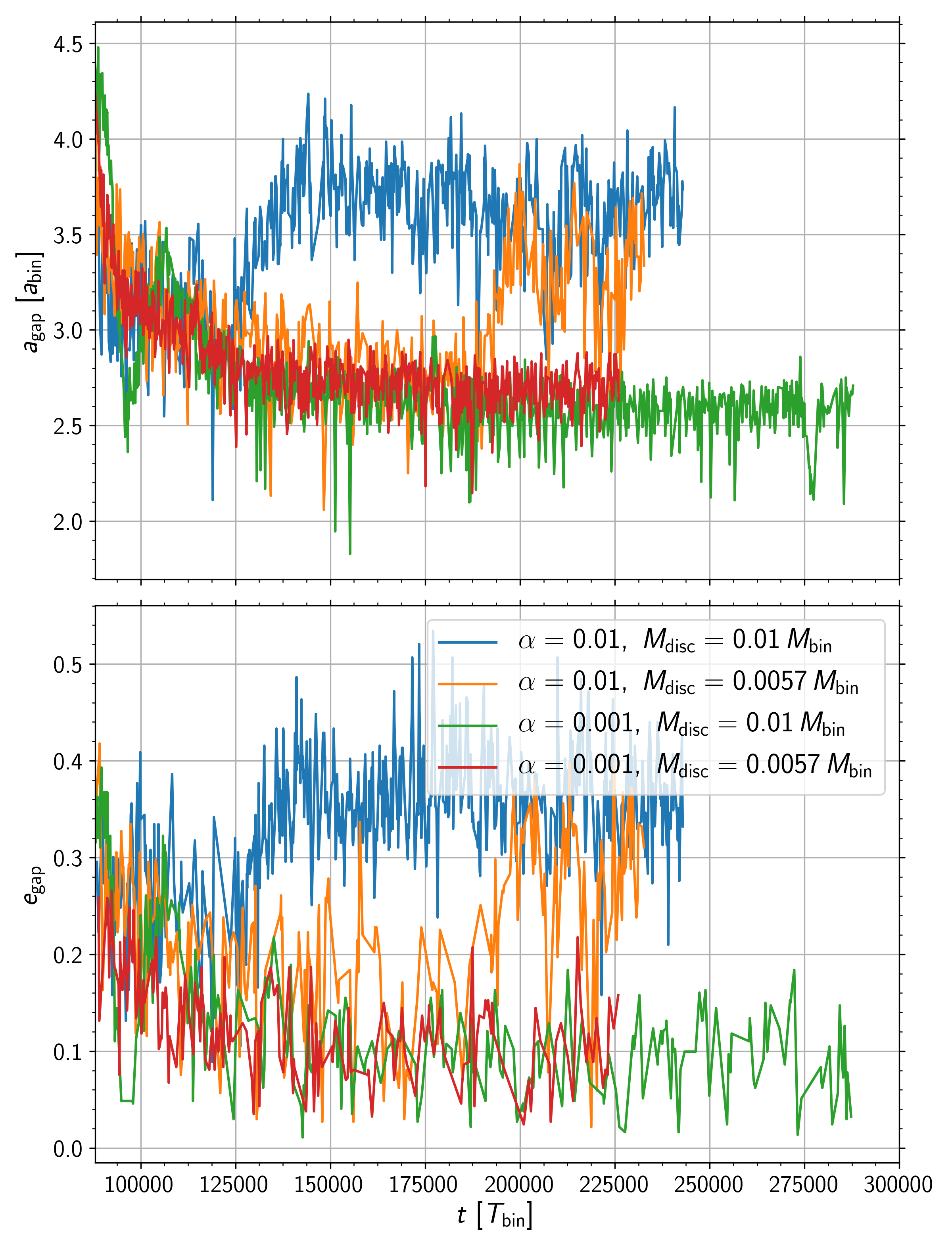}}
    \caption{The evolution of the semi-major axis and eccentricity of the disc's gap for the
    models shown in Fig.~\ref{img:Kep38-Planet-orbital_elements_ap60}.
    }
    \label{img:Kep38-disc_elements_with_planet}
\end{figure}

\section{Planet evolution}\label{sec:planet}
In addition to just studying the disc dynamics, we add now a planet to the last four simulations
of the Kepler-38 system and follow its evolution through the disc.
The planet, with the observed mass of Kepler-38b (see Tab.~\ref{tab:planetkepler}), is embedded after \num{88000} binary orbits into the system.
At this time the discs have approximately reached their equilibrium configurations, the corresponding 2D density distribution is
shown in Fig.~\ref{img:Kep38-2d_sigma_disc_physics}.
To reduce the simulation time, the planet's starting position lies close to the disc's inner edge at
$a_p(t=0) = 6 a_\mathrm{bin}$. As shown in previous simulations, the final 'parking slots' of planets
in circumbinary discs does not depend on their initial location \citep{2015A&A...581A..20K,2017MNRAS.469.4504M,2018A&A...616A..47T}.
After insertion, the planet is first held fixed on its initial circular orbit for 2000 $T_\mathrm{bin}$.
Starting at \num{90000} $T_\mathrm{bin}$, all three objects (binary stars and planet) feel the disc in these simulations and
the planets evolve their orbits under the gravitational torques from the disc.

Figure~\ref{img:Kep38-Planet-orbital_elements_ap60} shows the evolution of orbital
elements of the embedded planet for the four different cases.
As seen already in simulations of locally isothermal discs, the planets can reach two different final parking positions
\citep{2018A&A...616A..47T}. In the first case ('high state') the planets migrate first inward and then reverse to reach
a final position further away from the central binary with a relatively high eccentricity of
about $0.20$. In this state the orbit of the planet is fully aligned with the eccentric inner hole of the disc and precesses
with the same speed. Here, this configuration is reached for the two high viscosity models using $\alpha = 0.01$.
In the second situation for the low viscosity, the planets continue their inward migration and reach
a final position further in ('low state') with very small
orbital eccentricity. This final parking position lies very close to the observed location of the planet, and the low orbital eccentricity
agrees with the observation as well.

\begin{figure*}
    \centering
    \includegraphics[width=17cm]{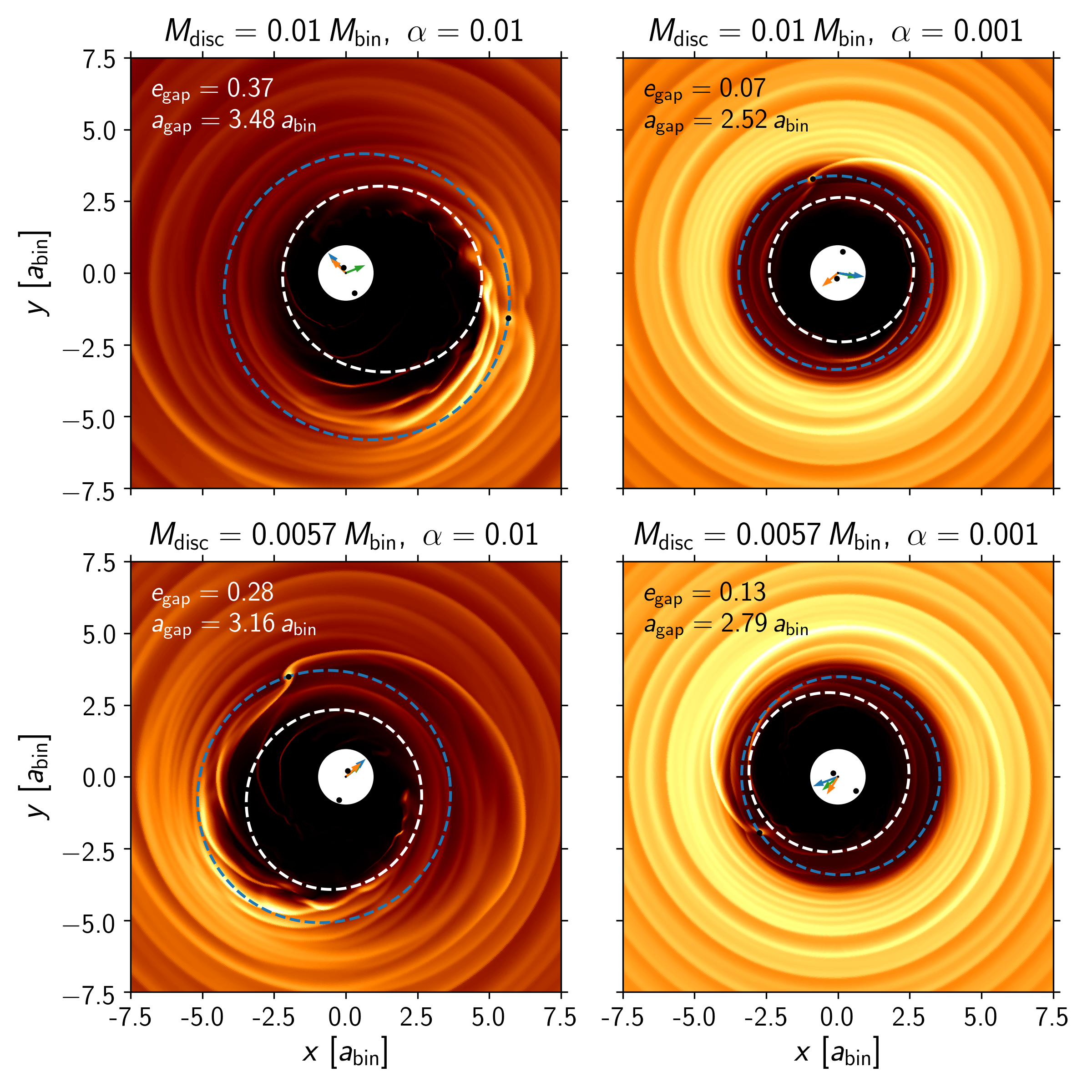}
    \caption{Structure of the inner disc for the Kepler-38 system with an
    embedded planet for different disc parameters. The surface
    density is colour-coded, since the absolute density values are not important
    for the disc structure we omit the density scale for clarity (brighter
    colours mean higher surface densities). The disc structure is shown after
    \num{225000} binary orbits. The binary
    components and the planets are marked with black dots. The
    direction of pericentre of the secondary (green), the planet (blue) and the disc (orange)
    are marked by the arrows. Additional fitted ellipses to the central
    cavity are plotted in dashed white lines and the actual orbit of the
    planet is shown as dashed blue line. Using the fitted ellipses, the size and
    eccentricity of the disc gap, shown in the upper left edge of each panel, are calculated at the same
    time.}
    \label{img:Kep38-2d_sigma_planets}
\end{figure*}

In \citet{2018A&A...616A..47T} we argued that it depends on the planet mass which state is reached eventually but there we only
performed models with one single (high) viscosity of $\alpha = 0.01$. Here, the planet mass is the same in all models, and the main difference is
the disc viscosity. The reason for the different behaviour of the planets lies in their ability to open a gap in the density of the disc.
In the lower viscosity models the planet mass of $0.38 M_\mathrm{Jup}$ is sufficient to open a partial gap in the surface density of the disc
while this is not the case for the higher viscosity. Once the planet induced gap is opened, the outer disc is separated from the inner one and
hydrodynamical waves cannot propagate through the opened gap. As a consequence, the binary is not able to excite eccentricity in the
outer disc, and it becomes more circular which is accompanied by a reduction in gap size.
This allows the planet to migrate further in and remain on a nearly circular orbit.
To corroborate this statement we ran an additional model using a locally isothermal disc with a small viscosity, $\alpha = 0.001$.
As expected, in this case the planet opens a partial gap and it continues to migrate towards a position very close to the observed
one with low eccentricity.

The strong mutual interaction of the disc with the planet is supported with
Fig.~\ref{img:Kep38-disc_elements_with_planet} where we show the time evolution of gap's semi-major axis and eccentricity.
Comparing to the planets' elements in Fig.~\ref{img:Kep38-Planet-orbital_elements_ap60} demonstrates that
the gap behaves exactly as the planets with respect to the increase and decrease of size and eccentricity.
The calculation of the gap's elements is quite noisy because the presence of the planet disturbs the density of the disc making
the automated fit of approximate ellipses \citep{2017A&A...604A.102T} more difficult.

The equilibrium density configuration of the disc is shown in Fig.~\ref{img:Kep38-2d_sigma_planets} after \num{225000} binary orbits together with
the positions of the binary stars and the planets. Additionally, we have indicated the direction of pericentres of binary, disc and planets,
with green, orange and blue arrows, respectively. The white ellipses denote the inner gap of the disc and the blue ones the orbit of the planet.
As seen already in the time evolution plots, in the low viscosity models (right panels) the planets have moved in closer and are on more circular orbits, and the discs
inner holes are also smaller and more circular. The planets orbit well inside the regions of maximum density of the disc.
In the case of high viscosity (left panels) the planets are more in touch with the disc and the density maxima sometimes lie inside of the planetary orbit.
In this case the planet is not able to clear its surroundings to form its own gap due to the high viscosity which tends to fill in the gaps.

\section{Discussion}\label{sec:discussion}
In this section we discuss several aspects concerning the physics of the radiative disc models and the planet migration process.
\subsection{Role of thermodynamics} \label{subsec:thermodynamics}
One goal of this work was to investigate what impact the disc's thermal structure has on its
dynamical evolution. For this purpose we extended our previous locally isothermal models
in \citet{2017A&A...604A.102T,2018A&A...616A..47T} taking viscous energy dissipation into account and assuming a cooling prescription
using a vertically averaged opacity. In such a situation the thermal structure of an accretion disc will be given
by the viscosity coefficient (here $\alpha$) as this determines the heating rate, and the mass contained in the disc as this
will determine the cooling rate through the density dependent opacity.
Neglecting for the moment compressive heating, e.g. by shocks, the temperature evolution follows from
\begin{equation}
 c_v \Sigma  \frac{\partial T}{\partial t}  = \Phi  - Q_-  \,,
\end{equation}
where $\Phi$ denotes the viscous dissipation, and $Q_-$ the cooling from
eq.~\eqref{eq:cooling_term}. Taking for a simple estimate the dominant $r\varphi$-component
of the stress tensor into account, the dissipation is given by
eq.~\eqref{eq:visc_dissipation}.
In the inner regions of the disc the opacity for our models varies as $\kappa \propto T$
and shows no density dependence, see Fig.~\ref{img:rosseland_mean_opacity}.
The viscosity varies as $\nu = \alpha c_s^2 / \Omega_K$ (see eq.~\eqref{eq:nu}) and the
sound speed is $\propto T^{1/2}$.
For the equilibrium temperature of the disc we then find, using $\Phi = Q_-$, the following dependence on mass and viscosity
\begin{equation}
    \label{eq:T_eq}
    T_\mathrm{eq}^2  \propto  \Sigma^2 \alpha \,.
\end{equation}
This relation explains very well the fact that the disc temperature in the four disc models
displayed in Fig.~\ref{img:Kep38-az_avg_disc_physics} are very similar to each other at the edge of the central cavity.
In the models with the reduced viscosity the density is higher because less mass is transferred into the cavity
which would result in a higher temperature but this is compensated by the lower viscosity.
In the outer regions of the discs the densities are similar and the temperature is reduced for the small viscosity models.
In addition to the viscous dissipation we checked the impact of shock heating, given by the work done by compression,
$-P  \, \nabla \cdot \vec{u}$.
We find that this term is highly variable and only contributes near the inner rim of
the central gap.   

\subsection{Comparison to locally isothermal models} \label{subsec:local_iso}
Concerning the overall evolution of the system, one important difference between our previous locally isothermal models
\citep{2018A&A...616A..47T} seems to be the extended initial phase before the system eventually reaches an equilibrium sate.
This is most likely caused by the fact that in the full models the initial cooling in regions with lower density results
in low temperatures with low pressure and small viscosity such that very large and highly eccentric cavities are produced in the early phase,
as shown for example in Fig.~\ref{img:Kep38-disc_props}. In the subsequent evolution the disc adjusts slowly on the longer
viscous timescale and closes the inner disc region again. In the locally isothermal models this effect was less pronounced
because the temperature, and consequently the viscosity, in the disc were always fixed
and the disc did much less show this overshoot behaviour 
especially for high $\alpha$ and/or high binary eccentricities.
The models for the Kepler-16 parameter show a very similar initial dynamical behaviour, as demonstrated
in Figs.~\ref{img:Kep16-disc_props_ecc} and \ref{img:Kep16-radial_plots_016}.
The transient feature seen here during the initial evolution can only be a feature of real discs
in case of a sudden change of the disc properties because slow variations would lead to a continuous re-adjustment.
Sudden changes of the disc could occur for example by a flyby of an additional object.

On the other hand, the final equilibrium states for both cases are quite similar, when locally isothermal models
are compared to radiative ones which show a similar disc height.
In our previous study \citep{2018A&A...616A..47T} we constructed sequences of locally isothermal models using a fixed $H/r=0.05$
and $\alpha =0.01$ for Kepler-38 parameter. In the situation studied here, the two radiative models using the smaller disc mass ($0.0057 M_\mathrm{bin}$)
result in temperatures that give a $H/r$ value very close 0.05 in the inner disc regions for both viscosities.
And indeed, the gap size and eccentricity are very compatible. In \citet{2018A&A...616A..47T} we find in equilibrium
for Kepler-38 values of $a_\mathrm{gap}=4.5 a_\mathrm{bin}$ and $e_\mathrm{gap} = 0.34$, while here we obtain
for the $\alpha = 0.01$ model  $a_\mathrm{gap}=4.04 a_\mathrm{bin}$ and $e_\mathrm{gap} = 0.37$ (orange curves in Fig.~\ref{img:Kep38-disc_props}).
From these viscous heated simulations we can see that in isothermal simulations a constant aspect ratio is reasonable for the inner part of a binary system
up to about $15-20 a_\mathrm{bin}$.

\subsection{Irradiation} \label{subsec:irradiation}
The models presented in this work do not include stellar irradiation and here we justify this simplification.
In Fig.~\ref{img:Kep38-az_avg_disc_physics} we overplot in the temperature panel the expected equilibrium temperature
of a disc that is heated by a single star with one solar mass.
This serves as a conservative estimate for the expected irradiation because it
is slightly more luminous than both binary stars together.
For the inner $10 a_\mathrm{bin}$ viscous heating is the primary heat source of the main disc.
Due to the very low density within the gap viscous heating is strongly reduced and irradiation becomes more important there.
Dynamically, the effects are limited because of the little mass contained in this region, but a more detailed treatment should take irradiation into account.

Beyond this radius irradiation can become more important, depending on viscosity and disc mass.
As the luminosity of a star scales with stellar mass to the power of about 3.5,
binary systems have a much lower luminosity for the same total mass than a single star system.
Therefore, in the case of a binary, viscous heating can become more relevant than irradiation.
In the outer regions of the disc irradiation, which scales with $L_*^{1/8}/M_*^{1/2}$,
becomes more important and it affects the value of the aspect ratio.

\subsection{The inner disc region} \label{subsec:inner_disc}
Using a polar grid does not allow to cover the central region which include the binary stars.
Based on our previous studies \citep{2017A&A...604A.102T}, where we tested in detail the impact of the location of the
inner boundary on the results, we performed all the present simulations with a value $R_\mathrm{min}= a_\mathrm{bin}$.
As all simulations show clear gaps with very low densities this value of $R_\mathrm{min}$ should be sufficient for an accurate
determination of the disc dynamics around binaries.
However, this does not allow to follow the flow of gas inside of the gap and calculate which star will accrete the material eventually.

The imposed outflow boundary at $R_\mathrm{min}$ does not allow for mass flowing back into the computational domain.
In reality, gas leaving through the inner boundary could come back into the domain being 'flung out' by one of the binary stars.
But recent simulations \citep{2019ApJ...871...84M} show that gas entering the gap is accreted onto the stars
in circumstellar discs and does not influence the ambient circumbinary disc.
Hence, as the inner boundary in our model lies well within the gap it does not influence
the dynamics of the circumbinary disc.

Recently, \citet{2019ApJ...875L..21M} presented simulations of the inner region near the central binary where they resolve
the individual circumstellar discs covering \num{5000} binary orbits.
In a comparison simulation using our standard setup with $R_\mathrm{min} = a_\mathrm{bin}$
we find very similar features in the inflow regime.   
This demonstrates that even in the early phases of the simulation the inner region inside a binary separation does not alter the disc behaviour
and the chosen boundary condition creates an environment for the disc similar to a simulation with the full domain.

\subsection{The evolution of the binary} \label{subsec:binary_evol}
In the simulations with the migrating planet the central binary is allowed to evolve as well due to the gravitational action of the
disc. The disc torque will lead to a shrinkage of the orbit accompanied by an increase in eccentricity.
Over the whole time span of the active simulations from \num{90000} to \num{225000} $T_\mathrm{bin}$ the binary semi-major axis changes
only very little. The largest reduction (2.5\%) is seen for the model with $\alpha =0.01, M_\mathrm{disc}=0.01 M_\mathrm{bin}$,
the smallest (1.2\%) for  $\alpha =0.001, M_\mathrm{disc}=0.0057 M_\mathrm{bin}$.
At the same time the highest increase (to 0.123) in $e_\mathrm{bin}$ is seen for the model with $\alpha =0.001, M_\mathrm{disc}=0.01 M_\mathrm{bin}$
the lowest one (to 0.103) for the $\alpha =0.01, M_\mathrm{disc}=0.01 M_\mathrm{bin}$ model.
These changes are so small that they have no significant impact on the evolution of the planet or disc structure,
such that in future simulations of this kind the evolution of the binary can be neglected.

The mentioned test simulations in \citet{2017A&A...604A.102T} and the comparison to \citet{2017MNRAS.466.1170M}
   and to \citet{2019ApJ...875L..21M} 
   give us some confidence that by leaving out the innermost region of the binary we do not loose significant dynamics.
   This can only be checked in simulations where the whole domain is covered. The question whether the binary experiences
   an orbit shrinkage (as due to the disc torques) or an expansion (for example due to direct accretion of high angular
   material from the disc) can only be studied by computations that cover the whole domain. 
   Recent examples that find some evidence for an expansion of the binary are the simulations by \citet{2019ApJ...871...84M}
   or \citet{2019ApJ...875...66M}. It remains to be seen if this persists for discs with full thermodynamics.
 
\subsection{The evolution of the planet} \label{subsec:planet_evol}
In agreement with previous studies \citep{2013A&A...556A.134P,2014A&A...564A..72K} we have shown that an embedded
planet migrates to the edge of the central cavity. Additionally, we confirm the existence of two final states.
In the first case ('high state') the planetary orbit has a larger eccentricity (usually around $e_\mathrm{p} \approx 0.2$ to $0.3$) and
is fully aligned with the eccentric hole of the disc, i.e. planet and gap precess at the same rate. In the second case ('low state')
the orbit of the planet remains circular and the final semi-major axis is smaller than in the first case.
Here, and in our previous study \citep{2018A&A...616A..47T} we demonstrate that the high state is taken for low planet masses and/or high
viscosity while the low state for higher planet masses and/or low viscosity. Hence, it is linked to the ability of gap formation
by the planet in the disc which depends on exactly these two parameter \citep{2006Icar..181..587C}. This connection became apparent
in the locally isothermal simulations in \citet{2018A&A...616A..47T} where a variation of planet mass has shown that
lower mass planets end up in the high state while high mass planets reach the low state.
Here, we have investigated additionally the viscosity dependence.

Once a planet near the inner edge is able to open its own gap the outer disc is shielded from the action of the binary,
becomes more circular and allows for a continuation of the inward migration. When the planet is not able to open
a gap, its motion is dominated by the disc and its orbit remains eccentric. This close connection between disc and planet dynamics is
illustrated by a direct comparison of Figs.~\ref{img:Kep38-Planet-orbital_elements_ap60} and \ref{img:Kep38-disc_elements_with_planet}.
We have illustrated this behaviour for the Kepler-38 system, where the planet is more massive but it has been seen for Kepler-16 as well,
for example in the simulations by \citet{2017MNRAS.469.4504M} who discussed the evolution of growing cores in discs with low viscosity $\alpha = 0.001$.

\section{Summary}\label{sec:summary}

We have performed 2D numerical simulations of viscous accretion discs around binary stars where we included viscous
heating and radiative cooling. In a first study we explored, using a given binary mass ratio of $q_\mathrm{bin} = 0.29$ (as in Kepler-16)
and an initial disc mass of $1/100 M_\mathrm{bin}$, the impact of the binary eccentricity on the disc dynamics.
In a second part we focused on the Kepler-38 system and calculated the disc structure for two different values of disc masses and viscosities.
For the Kepler-38 we then followed the evolution of an embedded planet for the four models.

We find that in all cases the discs evolve towards a quasi-stationary state with an eccentric cavity that shows a slow prograde precession.
The time for attaining this equilibrium can take over $~\num{100000}$ binary orbits, starting from an axisymmetric initial
configuration. In contrast to locally isothermal models, the inner discs cools off initially in the centre such that gaps of large size and high
eccentricity are formed because of the reduced pressure and viscosity (see Figs.~\ref{img:Kep16-disc_props_ecc} and \ref{img:Kep38-disc_props}).
During the subsequent evolution the disc readjusts and settles to an equilibrium state where the in the inner disc region
the vertical pressure scale height is about $H/R = 0.04$ to $0.05$, which depends on disc mass and viscosity.
Given that these values are not too different from locally isothermal simulations that typically use $H/R = 0.05$,
the gap structures are comparable concerning size and eccentricity. However, in the general case, the agreement between isothermal
and radiative models will depend on the used parameter of the system (disc viscosity and mass). In the end, only radiative models
will give the correct temperature in the disc, which allows to predict a realistic aspect ratio profile for further studies.

For the displayed model sequences we find a similar dependence of the gap precession rate on binary eccentricity as in locally isothermal models.
Models with circular binaries produce holes with the highest eccentricity. Upon increasing $e_\mathrm{bin}$ the gap size becomes smaller,
less eccentric, and the precession period reduces until at a critical $e_\mathrm{bin} = 0.16$ the trend reverses.
The gaps become larger and more eccentric with an increase of the precession period. Hence, overall this behaviour is very similar
to the locally isothermal models even with the same turning point \citep{2018A&A...616A..47T}.

The temperature maximum of the disc lies inside of the density peak
and coincides for the models with $e_\mathrm{bin} > 0.1$ with the maximum of the viscous dissipation rate,
which is here the main heating source of the disc. Shock heating plays a role only in the very inner regions of the disc
and contributes more for the binaries with small $e_\mathrm{bin}$.  
Concerning stellar irradiation, we find that for the parameters used here, it did not play a role in determining the energy balance of the
disc. However, for more massive (luminous) stars and larger binary separations, this additional energy source will play a role and needs to be included.

Concerning the evolution of embedded planets we find that the planets migrate inward through the disc due the gravitational torques acting on them.
They are parked near the inner edge of the disc because the positive density gradient acts as a planet trap \citep{2006ApJ...642..478M}.
We find evidence for two different dynamical states of the final position. In the 'high' state the planet resides at larger separations 
on an eccentric orbit which is precessing at the same rate as the gap region of the disc, i.e. the planetary orbit is always fully aligned with
the disc. In the 'low' state the planet remains on a circular orbit and is able to move closer to the central binary star. We have shown that it is
the ability of the planet to open its own gap in the disc which separates the two states. The high state occurs for lower mass planets,
or if the disc's viscosity is too high that the planet cannot open its gap, such that its dynamics is determined by the eccentric disc.
On the other hand, for higher planet masses or lower viscosity the planet opens a gap and shields the outer disc from the action of the binary.
Hence, the disc remains circular and the planet can move in closer to the star keeping its circular orbit.

Our simulations show that the final position of the modelled planet lies slightly outside of the observed location of the Kepler-38
planet even for the low state result. Considering that the observed planet is on a nearly circular orbit, we can conclude that in the
final phase of the planetary migration process the disc was in a low viscosity state. This applies most likely to the majority of the observed circumbinary planets as well,
because they reside on orbits with very low eccentricity close to the binary \citep{2018MNRAS.480.3800H}. 
This connection of final parking position of planets with the disc viscosity in circumbinary systems 
will allow to put constraints on the disc viscosity via observed planetary orbits.
Only in the systems Kepler-34 and Kepler-413
the planets show significant eccentricity and in both cases the numerical models show an even higher eccentricity of the final planetary orbit
\citep{2017MNRAS.469.4504M,2018A&A...616A..47T}. These systems need further investigation.

In a recent study investigating the stability of planetary orbits around binary stars it was argued that there is no pile-up of
planets in circumbinary discs \citep{2016ApJ...831...96L} as in more than half of the Kepler systems it would be possible to place
an additional planet inside of the observed one on a stable orbit \citep{2018ApJ...856..150Q}. In spite of this possibility of stable orbits,
our results suggest that it might be difficult to place migrating planets that close due to the large disc gap opened by the binary. 
In this context it will be useful to investigate the migration process of two planets in the disc and study their final locations and stability
\citep{2015A&A...581A..20K} and investigate the possibility of packed orbits around binaries \citep{2014MNRAS.437.3727K}.
In this context the recent new finding of a third planet in Kepler-47 circumbinary system \citep{2019AJ....157..174O} will be of importance.

\begin{acknowledgements}
We acknowledge fruitful discussions with Richard Nelson.
Daniel Thun and Anna Penzlin were funded by grant KL 650/26 of the German Research Foundation
(DFG). Most of the numerical simulations were performed on the bwForCluster
BinAC. The authors acknowledge support by the High Performance and
Cloud Computing Group at the Zentrum f\"ur Datenverarbeitung of the University
of T\"ubingen, the state of Baden-W\"urttemberg through bwHPC and the German
Research Foundation (DFG) through grant no INST 37/935-1 FUGG. All plots in
this paper were made with the Python library matplotlib \citep{Hunter:2007}.
\end{acknowledgements}

\bibliography{cb_cooling}
\bibliographystyle{aa}

\begin{appendix}
\section{Implementation of the cooling term}
\label{app:cooling}
    As mentioned in the numerics section we use a modified version of
    \textsc{Pluto} 4.2 which was ported to run entirely on Graphical Processor
    Units (GPUs). The original version of \textsc{Pluto} contains a very
    sophisticated cooling module \citep{2008A&A...488..429T} which we did not
    implemented in the GPU version. We followed a simpler approach which will be
    described in this Appendix.

    Pluto advances the hydrodynamical equations by one time step solving
    the following equations (we use the Runge-Kutta 2 time step algorithm):
    \begin{align}\label{eq:hydro_rk2}
        \vec{U}^* &= \vec{U}^n + \Delta t \mathcal{L}^n \\
        \vec{U}^{n+1} &= \frac{1}{2}
                         \left(\vec{U}^n + \vec{U}^* + \Delta t \mathcal{L}^*\right) \,.
    \end{align}
    with the vector of conserved variable $\vec{U} = (\Sigma, \Sigma u_R, \Sigma
    u_\varphi, \Sigma e)^\mathrm{T}$ and the \enquote{right-hand-side} operator
    $\mathcal{L}$ (for details see \citet{2007ApJS..170..228M}). During the
    calculation of the \enquote{right-hand-side} operator, the contribution of
    the cooling term $\Delta t \cdot Q_{-}$ is subtracted from the total energy of the
    gas.

\begin{figure}
        \centering
        \resizebox{\hsize}{!}{\includegraphics{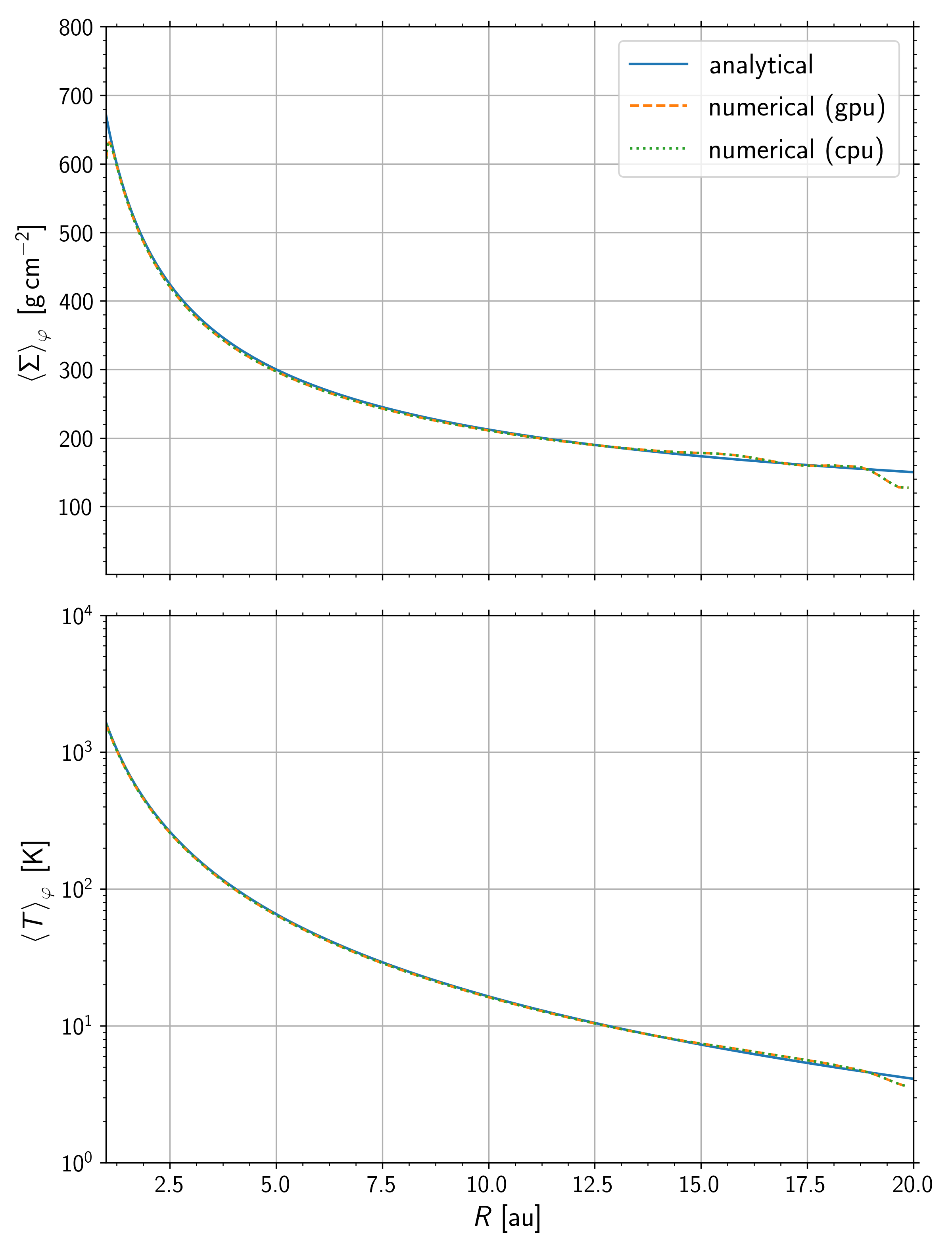}}
        \caption{The equilibrium structure of a disc around a single star used as a test problem
         for the implementation of the cooling term. Compared are the analytical solution
        according to eqs.~(\ref{eq:sig-ref}) and (\ref{eq:T-ref}) together with the results obtained using the CPU and GPU
        versions of the code.
        }
        \label{img:cooling_test}
    \end{figure}

    To test our simple implementation we used the following test problem
    by~\citet{2003ApJ...599..548D}, and simulate a Keplerian disc around a solar
    mass star with kinematic viscosity $\nu = \SI{5e16}{cm^2 s^{-1}}$. If we
    assume an optical thick disc $\tau \gg 1$ and an opacity given by
    \begin{equation}
        \kappa = \kappa_0 \, \left(\frac{T}{\mathrm{K}}\right)^2\,\mathrm{cm}^2\,\mathrm{g}^{-1}\,,
    \end{equation}
    with $\kappa_0 = \num{2e-6}$. The cooling term is given by
    \begin{equation}
        Q_- =  \frac{16}{3} \frac{\sigma_\mathrm{R} T^4}{\kappa \Sigma}
        \sqrt{2 \pi}  \,,
    \end{equation}
    see eqs.~\eqref{eq:cooling_term}, \eqref{eq:tau_eff} and
    \eqref{eq:optical_depth}.  In such a disc the equilibrium surface density
    and temperature are given by
    \begin{align}
        \Sigma &= 300 \sqrt{\frac{\SI{5}{au}}{R}}\,\frac{\mathrm{g}}{\mathrm{cm}^2} \label{eq:sig-ref} \,, \\
        T &= \frac{104}{(2\pi)^{1/4}} \left(\frac{\SI{5}{au}}{R}\right)^2\,\mathrm{K} \label{eq:T-ref} \,.
    \end{align}
    For details see \citet{2003ApJ...599..548D}. Figure~\ref{img:cooling_test} shows
    the numerical result from the CPU and GPU version of the code as well as the
    analytical solution. In our simulation we started with a constant surface
    density $\Sigma(t=0) = \SI{197}{g cm^{-2}}$ and a constant temperature
    of $T(t=0)=\SI{352}{K}$. One can see that both versions of the code
    reproduce the analytical equilibrium states of the surface density and
    temperature. Deviations at the inner and outer boundary come from the closed
    boundary conditions applied there.

\begin{figure}
        \centering
        \resizebox{\hsize}{!}{\includegraphics{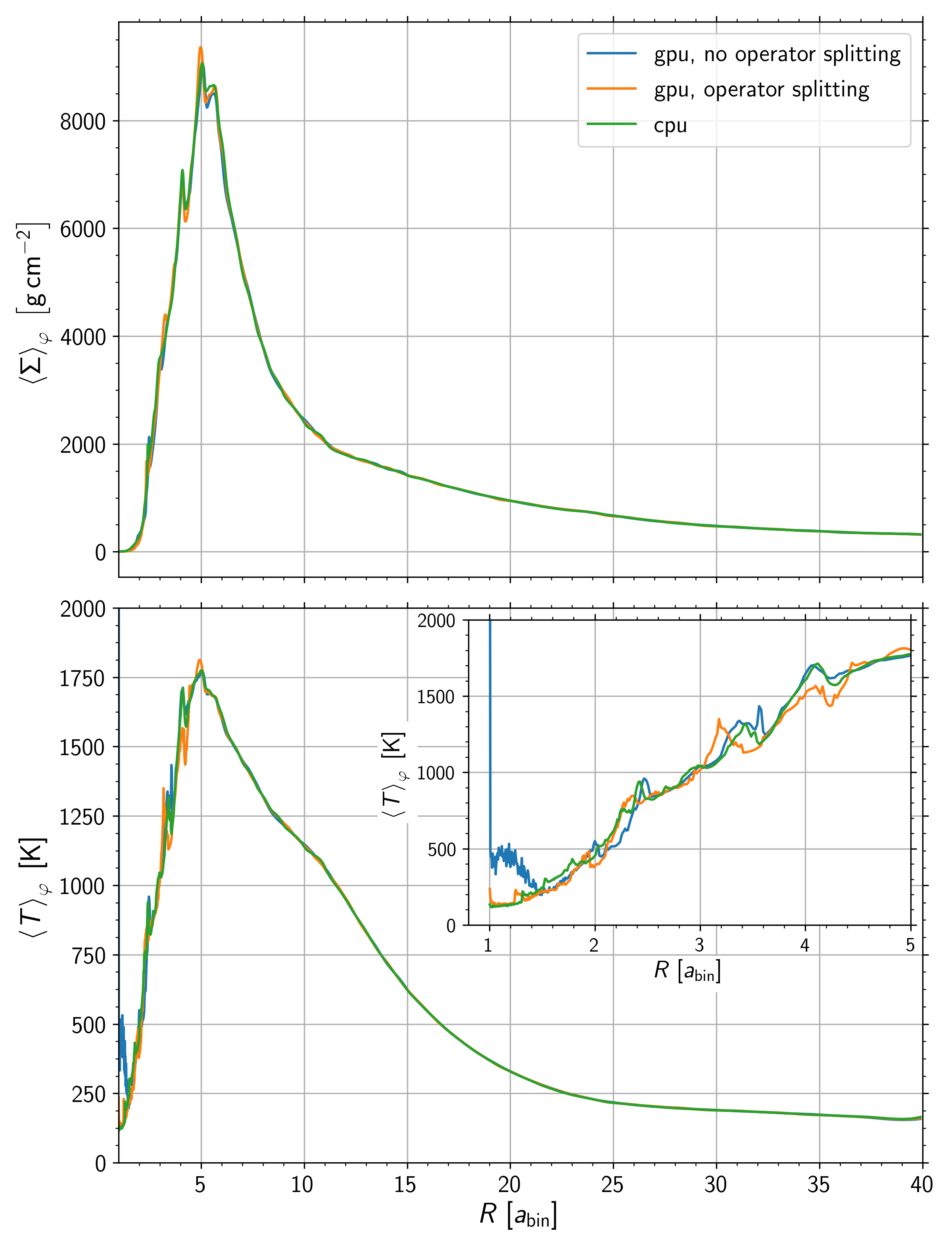}}
        \caption{A direct comparison of the disc structure around the
        binary star for the GPU and CPU versions of the code. The CPU version
        uses the implementation described in \citet{2008A&A...488..429T}. The
        GPU uses a simpler implementation described in this appendix. No
        operator splitting means that the cooling term is directly added during
        the hydro step, whereas operator splitting means that after the hydro
        step an additional differential equation involving only the cooling term
        is solved \eqref{eq:cooling_dgl} with a simple Euler method.  Shown is
        the surface density and temperature structure after \num{1500} binary
        orbits. The inlet in the bottom panel shows a zoom in at the inner
        boundary of the computational domain.}
        \label{img:cmp_gpu_cpu}
    \end{figure}

    During our circumbinary disc simulations we noticed that the
    temperature rises to very high values at the inner boundary of the computational domain, where the densities are very low.
   To examine this behaviour, we therefore computed models for the full setup,
    as described in Sec.~\ref{sec:setup}, using different methods of solving for the cooling.
    The surface density and the temperature are compared after \num{1500} binary orbits for the GPU
    and CPU version of the code.
    The results are summarised in Fig.~\ref{img:cmp_gpu_cpu}.
    The blue curves in Fig.~\ref{img:cmp_gpu_cpu} shows result from the GPU
    version with the cooling term implemented as described above. The inlet in
    the bottom panel shows the rise of the temperature at the inner boundary.
    The CPU version using the module from  \citet{2008A&A...488..429T} (green curve)
    does not show this rise of temperature. We
    therefore concluded that it is a numerical artefact.
    To test this hypothesis we changed the implementation of the cooling term in
    the GPU version from the direct approach above to an operator splitting method.
    In this new implementation the hydrodynamical equations~\eqref{eq:hydro_rk2}
    are solved first, without adding the cooling term. After the
    hydro step, the following equation is solved as a separate substep in the code.
    \begin{equation}\label{eq:cooling_dgl}
        \Prt{t}\vec{U} =  \vec{S}_\mathrm{cooling}
    \end{equation}
    with the simple Euler method
    \begin{equation}
        \vec{U}^{n+1} = \vec{U}^n + \Delta t \, \vec{S}_\mathrm{cooling}
    \end{equation}
    The cooling source term is $\vec{S}_\mathrm{cooling} = (0, 0, 0,
    - Q_{-})^\mathrm{T}$, with $Q_{-}$ given by eq.~\eqref{eq:cooling_term}.

    The orange curve in Fig.~\ref{img:cmp_gpu_cpu} shows result using this
    operator splitting technique. Here the temperature remains low at the inner
    boundary, comparable to the CPU version which also uses an operator
    splitting technique. The results shown in this paper are however calculated
    with the direct approach described at the beginning of this appendix. Since
    the surface density and pressure at the inner boundary are very low. The
    numerically increased temperatures do not alter our results, but in the
    future we will use the more suitable operator splitting method.

    The time step of the CPU version is more than two orders of magnitude
    smaller than the time step of the GPU version. This due to the complex cooling
    module implemented in the CPU version \citep{2008A&A...488..429T}. Our GPU
    version implements the cooling term in a much simpler way, as described in this
    appendix. Due to the very low time step of the CPU version we compared the
    results of the two codes already after \num{1500} binary orbits.
    In cases where the simple Euler step does lead to numerical problems it could be applied
    multiple times during one hydro time step or solved with an iterative method.
\end{appendix}

\end{document}